\documentclass[twocolumn,pra,aps,superscriptaddress,showpacs,floatfix]{revtex4}

\usepackage{graphicx}
\usepackage{amsmath}
\usepackage{amssymb}
\usepackage{bm}
\usepackage{array}
\usepackage{pdfpages}
\usepackage{natbib}
\usepackage{stmaryrd}
\usepackage{bbold}

\graphicspath{{./figures/}}
\newcommand{\norme}[1]{\left\Vert #1\right\Vert}
    
\begin{document}
\title{Spin nematic order in antiferromagnetic spinor condensates}
\author{T. Zibold}
\affiliation{Laboratoire Kastler Brossel, Coll\`ege de France, CNRS, ENS-PSL Research University, UPMC-Sorbonne Universit\'es, 11 place Marcelin Berthelot, 75005 Paris}
\author{V. Corre}
\affiliation{Laboratoire Kastler Brossel, Coll\`ege de France, CNRS, ENS-PSL Research University, UPMC-Sorbonne Universit\'es, 11 place Marcelin Berthelot, 75005 Paris}
\author{C. Frapolli}
\affiliation{Laboratoire Kastler Brossel, Coll\`ege de France, CNRS, ENS-PSL Research University, UPMC-Sorbonne Universit\'es, 11 place Marcelin Berthelot, 75005 Paris}
\author{A. Invernizzi}
\affiliation{Laboratoire Kastler Brossel, Coll\`ege de France, CNRS, ENS-PSL Research University, UPMC-Sorbonne Universit\'es, 11 place Marcelin Berthelot, 75005 Paris}
\author{J. Dalibard} 
\affiliation{Laboratoire Kastler Brossel, Coll\`ege de France, CNRS, ENS-PSL Research University, UPMC-Sorbonne Universit\'es, 11 place Marcelin Berthelot, 75005 Paris}
\author{F. Gerbier\email{fabrice.gerbier@lkb.ens.fr}} 
\affiliation{Laboratoire Kastler Brossel, Coll\`ege de France, CNRS, ENS-PSL Research University, UPMC-Sorbonne Universit\'es, 11 place Marcelin Berthelot, 75005 Paris}
\date{\today}

\begin{abstract}
Large spin systems can exhibit unconventional types of magnetic ordering different from the ferromagnetic or N\'eel-like antiferromagnetic order commonly found in spin 1/2 systems. Spin-nematic phases, for instance, do not break time-reversal invariance and their magnetic order parameter is characterized by a second rank tensor with the symmetry of an ellipsoid. Here we show direct experimental evidence for spin-nematic ordering in a spin-1 Bose-Einstein condensate of sodium atoms with antiferromagnetic interactions. In a mean field description this order is enforced by locking the relative phase between spin components. We reveal this mechanism by studying the spin noise after a spin rotation, which is shown to contain information hidden when looking only at averages. The method should be applicable to high spin systems in order to reveal complex magnetic phases.
\end{abstract}
\maketitle

\section{Introduction}

Magnetic order in spin $1/2$ systems is commonly associated with either a ferromagnetic phase or a N{\'e}el antiferromagnet, depending on the sign of the exchange interactions. The situation is richer for spins greater than $1/2$, and other types of magnetic order can arise at low temperatures. Spin 1 systems, for instance, can support spin nematic phases with vanishing average spin $\langle \hat{\bm{s}} \rangle$ \cite{andreev1984a}. The magnetic order is then characterized by a non-zero spin quadrupole tensor, $Q_{ij} \equiv \frac{1}{2} \langle \hat{s}_i \hat{s}_j +\hat{s}_j \hat{s}_i \rangle$ which deviates from isotropy even without applied field, {\it i.e.} it describes an object with the symmetries of an ellipsoid. In the simplest case, with axial symmetry, the spin quadrupole tensor has the same mathematical form as the orientational order parameter of nematic liquid crystals \cite{degennes}. There is a preferred axis in space (the \textit{director}) without a preferred direction along that axis. 

Spin nematic phases have been identified in lattice spin 1 models (see, {\it e.g}, \cite{blume1969a,chen1971a,Nakatsuji2005a,podolski2005a,tsunetsugu2006a,bhattacharjee2006a,lauchli2006a,michaud2011a}) or in spin 1 Bose-Einstein condensates (BECs) \cite{stamperkurn2013a} with antiferromagnetic spin-exchange interactions \cite{stenger1998a,ohmi1998a,zhou2003a,imambekov2003a,zhou2004a,black2007a,liu2009a,bookjans2011a,jacob2012a,deforges2014a}. In solid state systems, most magnetic probes couple only to the magnetization and are therefore unsuitable to reveal spin nematic order. In spin 1 condensates, equilibrium properties have been characterized by measuring the populations of each Zeeman state. This is not always sufficient to establish the nature of the magnetic order. For instance, in the so-called broken axisymmetry phase \cite{ueda2012a}, where all three Zeeman sublevels are populated, ferromagnetic or spin nematic behavior cannot be distinguished from the average populations alone. 

\begin{figure}[t]
  \centering
    \includegraphics[width=0.48\textwidth]{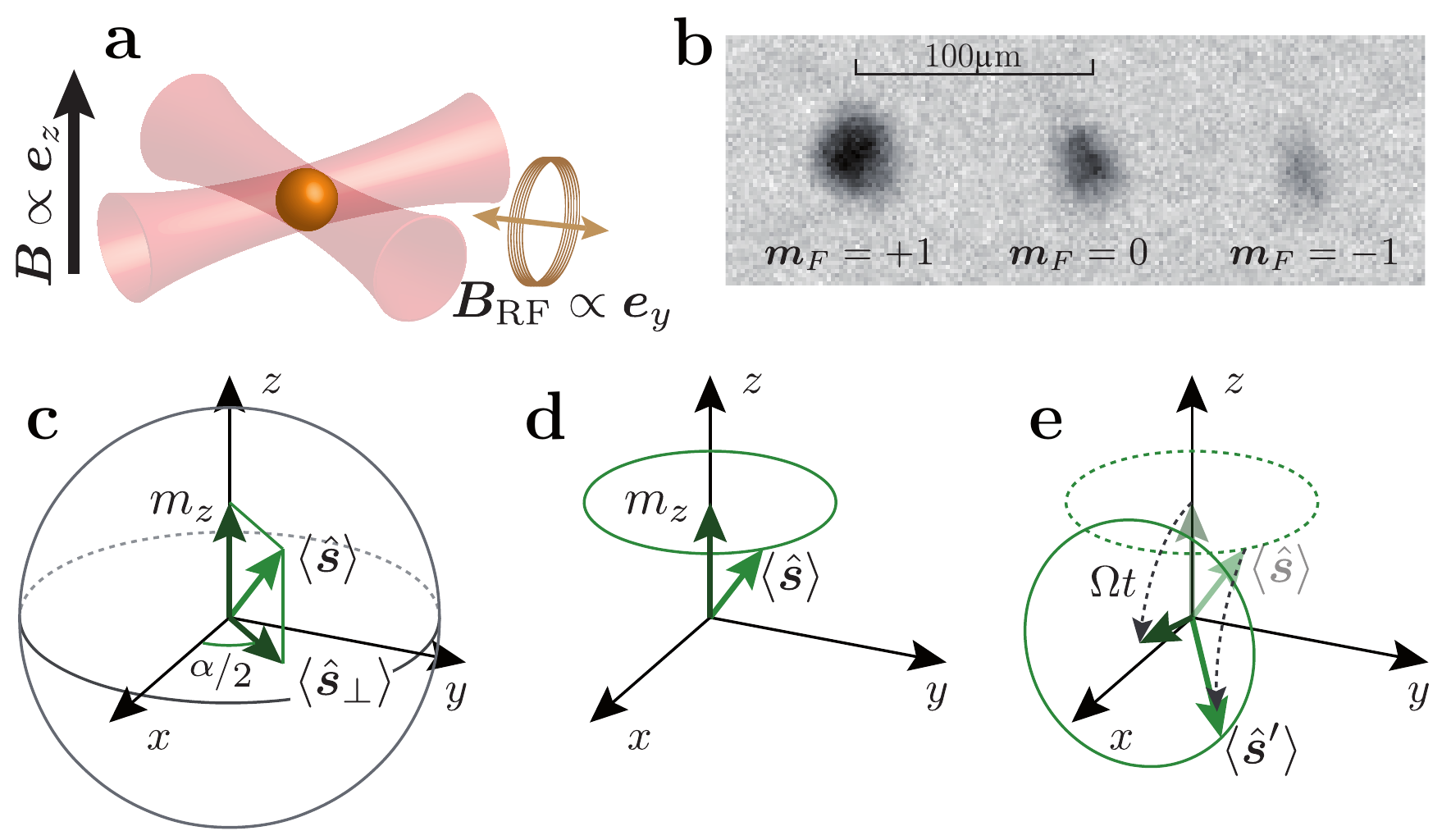}
    \caption{(Color online): {\bf (a)}: Sketch of the experimental setup. 7500 Bose-condensed $^\text{23}\text{Na}$ atoms are confined in a crossed optical dipole trap with a homogeneous static magnetic field along $z$. A resonant oscillating magnetic field along $y$ drives a spin rotation of the initial equilibrium state. {\bf (b)}: Absorption image of the atomic cloud after Stern-Gerlach expansion in a magnetic field gradient. {\bf (c)-(e)}: Classical picture explaining the principle of our measurement. {\bf (c)}: The average spin $\langle\hat{\bm{s}}\rangle$ of the condensate created in a single realization can be decomposed into a longitudinal component $m_z=\langle \hat{s}_z\rangle$ and a transverse component $\langle \hat{\bm{s}}_\perp \rangle =\langle \hat{s}_x \rangle \bm{e}_x+\langle \hat{s}_y \rangle \bm{e}_y$, the direction of which is given by the angle $\alpha$. {\bf (d)}: From realization to realization, the angle $\alpha$ varies randomly while $m_z$ and $\vert\langle \hat{\bm{s}}_\perp\rangle \vert$ stay constant. The mean spin vector $\langle\hat{\bm{s}}\rangle$ thus samples a horizontal circle of radius $\vert\langle\hat{\bm{s}}_\perp \rangle\vert$. {\bf (e)}: A spin rotation of the initial state rotates this circle by an angle $\Omega t$ along the $y$ axis. The fluctuations $\Delta s'_z$ after rotation are proportional to the squared radius of the circle through a simple geometrical relation.}
    \label{Fig1}
\end{figure}

In this article, we propose a method to reveal spin-nematic ordering (or possibly other types of unconventional magnetic order), and apply it experimentally to spin 1 atomic condensates. We show that the spin noise following a spin rotation contains information about the initial state, which can be retrieved with a suitable statistical analysis. In spinor condensates, magnetic order follows from the emergence of a well-defined phase relation between the components of the spin wavefunction in the equilibrium state. This phase-locking mechanism is not caused by any external field, but emerges from the interactions between the atomic spins. We show evidence for such a mechanism in a condensate of spin 1 $^{23}$Na atoms.

The article is organized as follows. In Section\,\ref{sec:theory} we recall results on the geometry of spin 1 wavefunctions, which are used to give a quantitative definition of spin nematic order. We connect it to the standard treatment of spinor condensates at $T=0$, and discuss the effect of finite temperatures. In Section\,\ref{sec:spinnoise}, we describe the method used to extract informations about the magnetic order from a measurement of spin noise after a known spin rotation. In Section\,\ref{sec:exp}, we describe our experimental apparatus and methods. Section\,\ref{sec:spinnoisemeasurement} describes a first analysis of our experimental results, where the fluctuations of magnetization after spin rotation are monitored. In Section\,\ref{sec:MLE}, another, more refined analysis is presented, where a maximum-likelihood estimation of the equilibrium single particle density matrix is presented. Both methods reveal the underlying spin nematic character of the equilibrium state. Section\,\ref{sec:conclusion} summarizes our findings.

\section{Theoretical description of antiferromagnetic spinor condensates}
\label{sec:theory}
The purposes of this Section are first, to give a precise definition of spin nematic phases in terms of spin observables, and second, to connect this definition to experiments with spin 1 Bose-Einstein condensates at $T=0$ and at finite temperatures. We will assume here that the spin 1 bosons are confined in a state-independent trap, tight enough to prevent the formation of spin domains in the equilibrium state (single-mode approximation) \cite{yi2002a}. The condensate wavefunction is then given by the product of a spatial mode function $\overline{\phi}(\bm{r})$, common to all Zeeman states, with a spin 1 wavefunction $\vert \bm{\zeta}\rangle$, which describes the internal degrees of freedom. An important feature of ultracold spinor gases is that the reduced (longitudinal) magnetization, $m_z=n_{+1}-n_{-1}$, is conserved by binary collisions driving the system to its equilibrium state \cite{stamperkurn2013a,jacob2012a}. Experimentally, we prepare a spin mixture well before the BEC forms in our evaporation sequence, allowing us to adjust the longitudinal magnetization $m_z$ between 0 and 1 (see Section\,\ref{sec:exp}). 

\subsection{Geometric description of spin 1 wavefunctions}

We first give a more precise definition of spin nematic order, and connect this definition with spin observables. To that end, it is convenient to express a spin 1 state in terms of its components in the so-called Cartesian basis $\{ \vert x \rangle,\vert y \rangle,\vert z \rangle\}$ \footnote{The Cartesian basis is defined as $\vert x \rangle = \frac{1}{\sqrt{2}}\left( \vert -1\rangle -\vert +1 \rangle \right)$, $\vert y \rangle =  \frac{i}{\sqrt{2}}\left( \vert -1\rangle +\vert +1 \rangle \right)$,
and $\vert z \rangle =  \vert 0 \rangle$. From the relation ${\hat S}_a \vert b \rangle =  i \epsilon_{abc} \vert c \rangle$ ($\epsilon_{abc}$ is the fully antisymmetric tensor), we deduce that the cartesian state $\vert a \rangle$ is the eigenstate of ${\hat S}_a$ with eigenvalue $0$.} formed by the eigenstates of ${\hat S}_a$ with eigenvalue $0$, where $a=x,y,z$. In this Section, we restrict ourselves to the case of pure states for simplicity. 

A spin 1 state can be written in the Cartesian basis as \cite{mullin1966a,ivanov2003a,zhou2004a,lauchli2006a}
\begin{align}
\vert \Psi \rangle = 
({\bm u}+i{\bm v}) \cdot \vert {\bm r}\rangle,
\end{align}
where the vectors ${\bm u}$, ${\bm v}$ are real and obey ${\bm u}^2+{\bm v}^2=1$. The vectors ${\bm u}$ and ${\bm v}$ are not uniquely defined. Performing a gauge transformation $\Psi \rightarrow \Psi'=e^{i\gamma} \Psi$ transforms ${\bm u}$ and ${\bm v}$ as ${\bm u}' = \cos(\gamma) {\bm u}- \sin(\gamma) {\bm v}$ and  ${\bm v}'=\cos(\gamma) {\bm v}+\sin(\gamma) {\bm u}$. As a result, we can choose $\gamma$ such that ${\bm u} \cdot {\bm v}=0$ and $\norme{{\bm u}} \geq \norme{{\bm v}}$.
 
The state of a spin 1 particle can be uniquely described by the average spin vector, $\langle \hat{\bm{s}} \rangle  =  2 \bm{u} \times \bm{v}$, and by the spin quadrupole tensor $Q_{ij} \equiv \frac{1}{2}\langle \hat{s}_i \hat{s}_j +\hat{s}_j \hat{s}_i \rangle $ ($\text{Tr}\overline{\overline{Q}}=2$). In the cartesian basis, we have
\begin{align}
Q_{ij}&  =  \delta_{ij} -(u_{i} u_{j}+v_{i} v_{j}),
\end{align}
or in a more geometrical form,
\begin{align}
\overline{\overline{Q}} &=\frac{1-\mathcal{A}}{2} \underline{\bm{u}} \otimes \underline{\bm{u}}+\frac{1+\mathcal{A}}{2} \underline{\bm{v}} \otimes \underline{\bm{v}}+\underline{\bm{w}} \otimes \underline{\bm{w}}.
\end{align}
The orthogonal units vectors $\underline{\bm{u}} = \bm{u}/\norme{\bm u}$, $\underline{\bm{v} } = \bm{v}/\norme{\bm v}$, $\underline{\bm{w}} = \langle \hat{\bm{s}} \rangle /\norme{\langle \hat{\bm{s}} \rangle}$ define the eigenaxis of $\overline{\overline{Q}}$, with eigenvalues $(1-\mathcal{A})/2$,$(1+\mathcal{A})/2$ and $1$. The {\it alignment parameter} $\mathcal{A}\geq 0$, defined as $\mathcal{A} = 2 \norme{\bm u}^2 -1$, characterizes the anisotropy of spin fluctuations in the plane perpendicular to the mean spin vector.


There are two simple limiting cases. The first one is the case of an aligned state (also called spin nematic or polar state in the context of spinor condensates \cite{stamperkurn2013a}), where the spin wavefunction,  $\vert \Psi \rangle = \underline{{\bm u}}\cdot \vert {\bm r}\rangle$, is the eigenstate of $\hat{\bm{s}}\cdot \bm{u}$ with eigenvalue zero. In such a state, the average spin vanishes, $\norme{\langle \hat{\bm{s}} \rangle}=0$, and the spin quadrupole tensor is $\overline{\overline{Q}} =\mathbb{1}- \underline{\bm{u}} \otimes \underline{\bm{u}}$ with eigenvalues $0,1,1$. In the literature, it is common to call $\bm{u} $ the director field. The tensor $\overline{\overline{Q}} $, or equivalently the director $\underline{\bm{u}}$, plays the role of the order parameter for spin nematic states.  

The second limiting case is the one of an oriented or fully magnetized state, for which the average spin is maximal, $\norme{\langle \hat{\bm{s}} \rangle}=1$. This is achieved when $\norme{{\bm u}}=\norme{{\bm v}}=1/\sqrt{2}$, and also corresponds to a non-zero spin quadrupole tensor
$\overline{\overline{Q}} =\frac{1}{2} \left( \mathbb{1}- \underline{\bm{w}} \otimes \underline{\bm{w}}\right)$ with eigenvalues $1/2,1/2,1$.

For a generic, partially magnetized state, one can quantify the proximity to one or the other limiting cases by the quantity $\mathcal{A}$, which characterizes the amount of alignment present in a given state. For purely aligned states $\mathcal{A}=1$ while for purely oriented states $\mathcal{A}=0$. For a generic state, the alignment $\mathcal{A}$ and spin length $\norme{\langle \hat{\bm{s}}\rangle}$ are related by
 \begin{align}\label{eq:A}
\langle \hat{\bm{s}} \rangle^2+{\mathcal{A}}^2=1.
\end{align}
This shows that measuring the length of the mean spin vector $\langle \hat{\bm{s}} \rangle ^2$ is fully equivalent to measuring the alignment $\mathcal{A}$. 


\begin{figure}
\begin{center}
\includegraphics[width=0.5\textwidth]{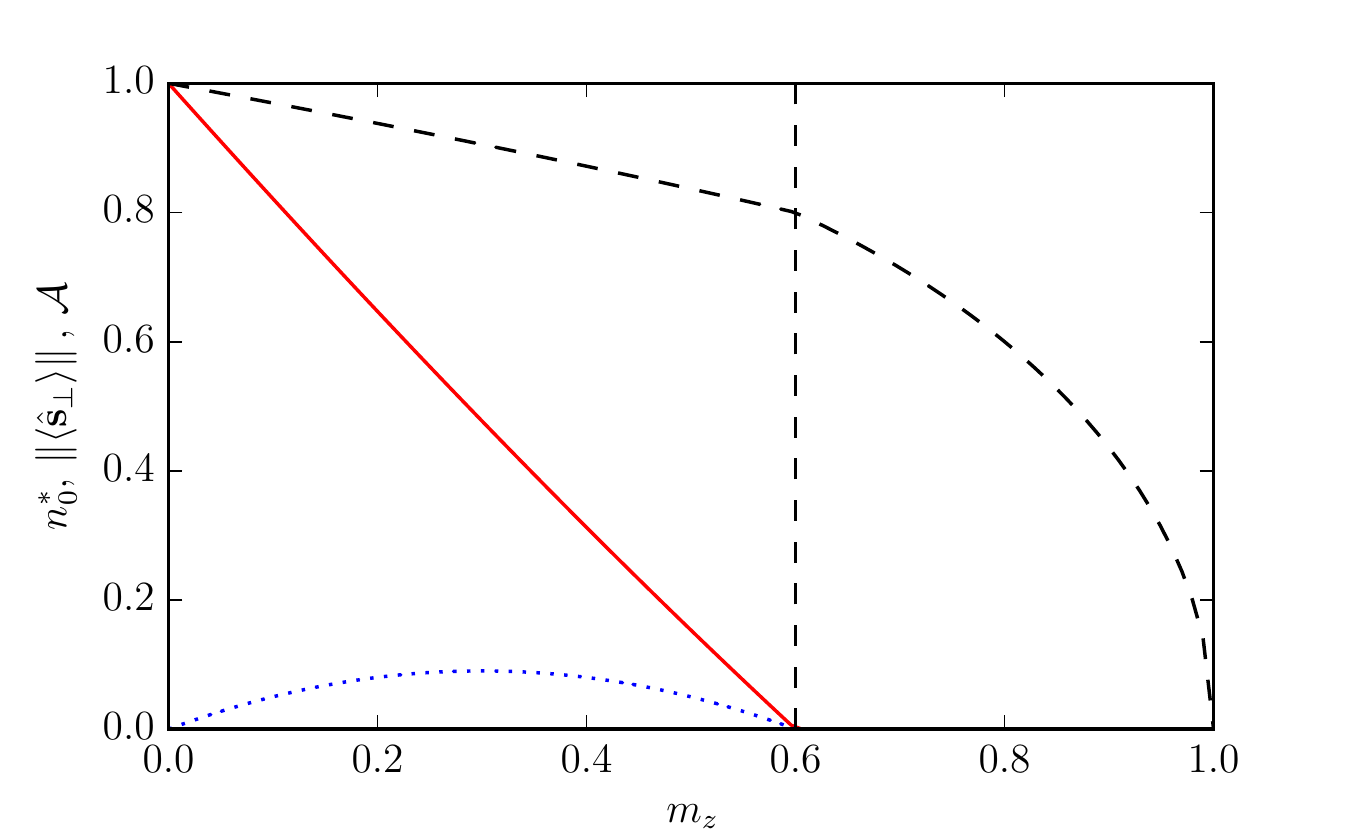}
 \centering 

 \caption{(Color online) Equilibrium population $n_0^\ast$ (solid line), transverse spin length $\norme{\langle\hat{\bm{s}}_\perp\rangle}$ (dotted line) and alignment $\mathcal{A}$ (dashed line) of an antiferromagnetic \mbox{spin 1} condensate versus longitudinal magnetization $m_z$, for a fixed value of $q/U_S=0.2$ (solid lines). The critical magnetization separing the broken axisymmetry from the antiferromagnetic phase is $m_{z,c}=0.6$, marked by the vertical dashed line. 
}
\label{fig:nematicNM}
\end{center}
\end{figure} 

\subsection{Ground state of spinor condensates} 
\label{sec:equilibrium}

In the single mode approximation where atoms in different spin states share the same spatial mode \cite{yi2002a}, we parametrize the spin state of the condensate as
\begin{align}
\label{eq:Psi}
\vert \bm{\zeta} \rangle
= \begin{pmatrix}
\sqrt{\frac{1-n_0+m_z}{2}}e^{i(\Theta+\alpha)/2}\\
\sqrt{n_0}\\
\sqrt{\frac{1-n_0-m_z}{2}}e^{i(\Theta-\alpha)/2}
\end{pmatrix},
\end{align}
where $\Theta$ and $\alpha$ are relative phases \footnote{The full Hilbert space can be parametrized by $\alpha \in [0,4\pi[$, $\Theta \in [0,2\pi[$, $n_0\in [0,1]$ and $\vert m_z\vert \leq 1-n_0$.}. The quantum state $\vert\bm{\zeta}\rangle$ corresponds to a mean spin vector $\langle \hat{\bm{s}} \rangle=m_z \bm{e}_z+\langle\hat{\bm{s}}_\perp\rangle$ (quantities in small letters are normalized by the total atom number $N$). The mean transverse spin $\langle\hat{\bm{s}}_\perp\rangle =\langle \hat{s}_x \rangle \bm{e}_x+\langle \hat{s}_y \rangle \bm{e}_y$ points in a direction determined by $\alpha$ and its length is determined by $\Theta$,
\begin{align}
\label{eq:Sperp}
\langle\hat{\bm{s}}_\perp\rangle^2 &=2n_0 \left(1-n_0+\sqrt{(1-n_0)^2-m_z^2} \,\cos\Theta \right).								
\end{align}
Eq.\,(\ref{eq:A}) shows that the relative phase $\Theta$ also determines the alignment of the state $\vert \bm{\zeta} \rangle$.

For a given magnetization $m_z$ set by the preparation sequence, the equilibrium state $\vert\bm{\zeta}\rangle$ minimizes the spin mean field energy $E_{\rm MF}$, the sum of the spin-exchange interaction energy and of the quadratic Zeeman energy (QZE) energy in an applied magnetic field $\bm{B}$ \cite{stamperkurn2013a},
\begin{align}\label{eqMFEnergy}
\frac{E_\text{MF}}{N}  = \frac{U_s}{2} \langle\hat{\bm{s}}_\perp\rangle^2-q  n_0,
\end{align}
up to terms that depend only on $m_z$. For the experiments reported in this article, the interaction strength is $U_s/h\approx 38\,$Hz (see Section\,\ref{sec:Us}) and $q/h\approx4\,$Hz to $34\,$Hz. 

Antiferromagnetic interactions ($U_s>0$, the case of sodium atoms) favor minimizing the transverse spin length. According to Eq.\,(\ref{eq:Sperp}), this is achieved by locking the relative phase $\Theta$ to $\pi$ independently of the value taken by $n_0,m_z,\alpha$ (ferromagnetic interactions would lock $\Theta$ to 0 instead). This is equivalent to maximizing the alignment $\mathcal{A}$ introduced above.

\begin{figure}[t]
  
  \centering
    \includegraphics[width=0.5\textwidth]{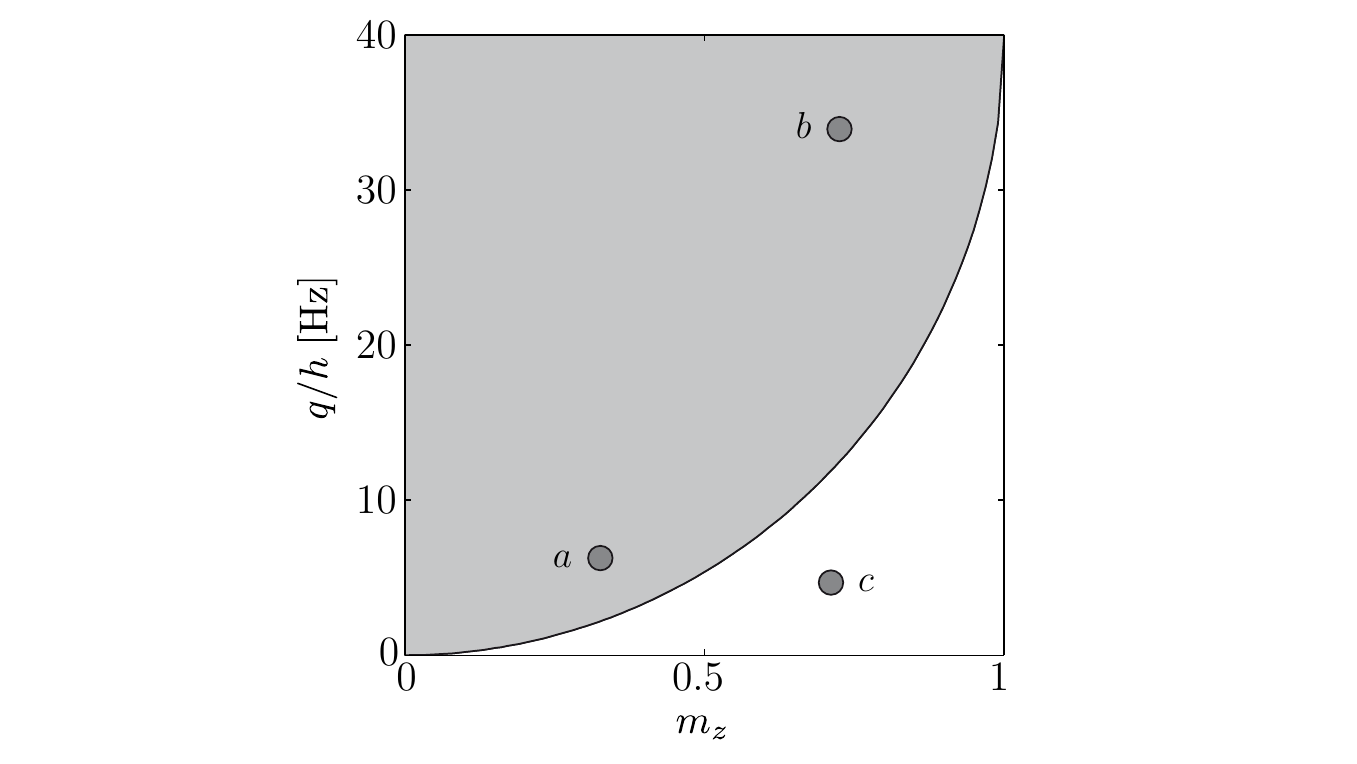}
    \caption{(Color online) The phase diagram in the $m_z-q$ plane, where the three sets of experiments we have performed are located, denoted as \textit{a},\textit{b},\textit{c} (\textit{a}: $\overline{m}_z=0.33$, $q/h=6.00$ Hz; \textit{b}: $\overline{m}_z=0.73$, $q/h= 33.7$ Hz; \textit{c}: $\overline{m}_z=0.71$, $q/h= 3.84$ Hz;). In the gray area above the phase transition line both $n_0$ and $\vert \langle\hat{\bm{s}}_\perp\rangle\vert$ are nonzero, whereas both vanish below the phase transition in the zero temperature case.}
    \label{Fig5}
\end{figure}

For a partially magnetized system with given magnetization $m_z$, the competition between the two terms in Eq.\,(\ref{eqMFEnergy}) drives a phase transition at a critical $q_c = U_s ( 1-\sqrt{1-m_z^2} )$ \cite{stenger1998a,zhang2003a,black2007a,liu2009a,jacob2012a}. At zero temperature, the equilibrium population $n^*_0$ is zero below $q_c$ (``antiferromagnetic phase'') and assumes a finite value above (``broken axisymmetry phase'') \cite{stenger1998a,zhang2003a,ueda2012a}, as illustrated in Fig.\,\ref{Fig5}. Fig.\,\ref{fig:nematicNM} shows the equilibrium population $n^*_0$, together with the length $\norme{\langle \hat{\bm{s}}_\perp\rangle}$ of the transverse spin and the alignment $\mathcal{A}$. Although the mean transverse spin is not zero above $q_c$ [see Eq.\,(\ref{eq:Sperp})], its value remains small because $\Theta$ stays locked to $\pi$. As a result, the alignment
\begin{align}
\mathcal{A} & =n_0 + \sqrt{(1-n_0)^2-m_z^2},
\end{align}  
which would reach $1$ in the absence of other constraints (thus realizing pure spin nematic states), stays very close to the maximum value given the conservation of $m_z$, $\mathcal{A}_\text{max} = \sqrt{1-m_z^2}$. This justifies using the transverse spin length to determine the amount of alignment present in the state $\vert \bm{\zeta} \rangle$, even when $\langle\hat{\bm{s}}\rangle \neq 0$.

\subsection{Finite temperatures}\label{sec:finiteT}

At finite temperatures, the $T=0$ description of a spinor condensate should be modified in two ways. First, the spin state of the condensate is subject to thermal fluctuations, and second, the population of the condensate is thermally depleted. In this Section, we examine these two effects in order. 

We first discuss the thermal fluctuations of the spin state of the condensate, which is described by a finite temperature spin ensemble as studied in details in \cite{corre2015a}. Close to the phase transition at $q_c$, the population $n_0^\ast >0$ which minimizes the free energy is small. The spin state of the condensate is then well described by a statistical mixture of $\vert N:\bm{\zeta}\rangle$ states, with an approximately Gaussian distribution of $n_0, m_z,\Theta$ \cite{corre2015a}. 


\begin{figure}
\centering
\includegraphics[width=0.5\textwidth]{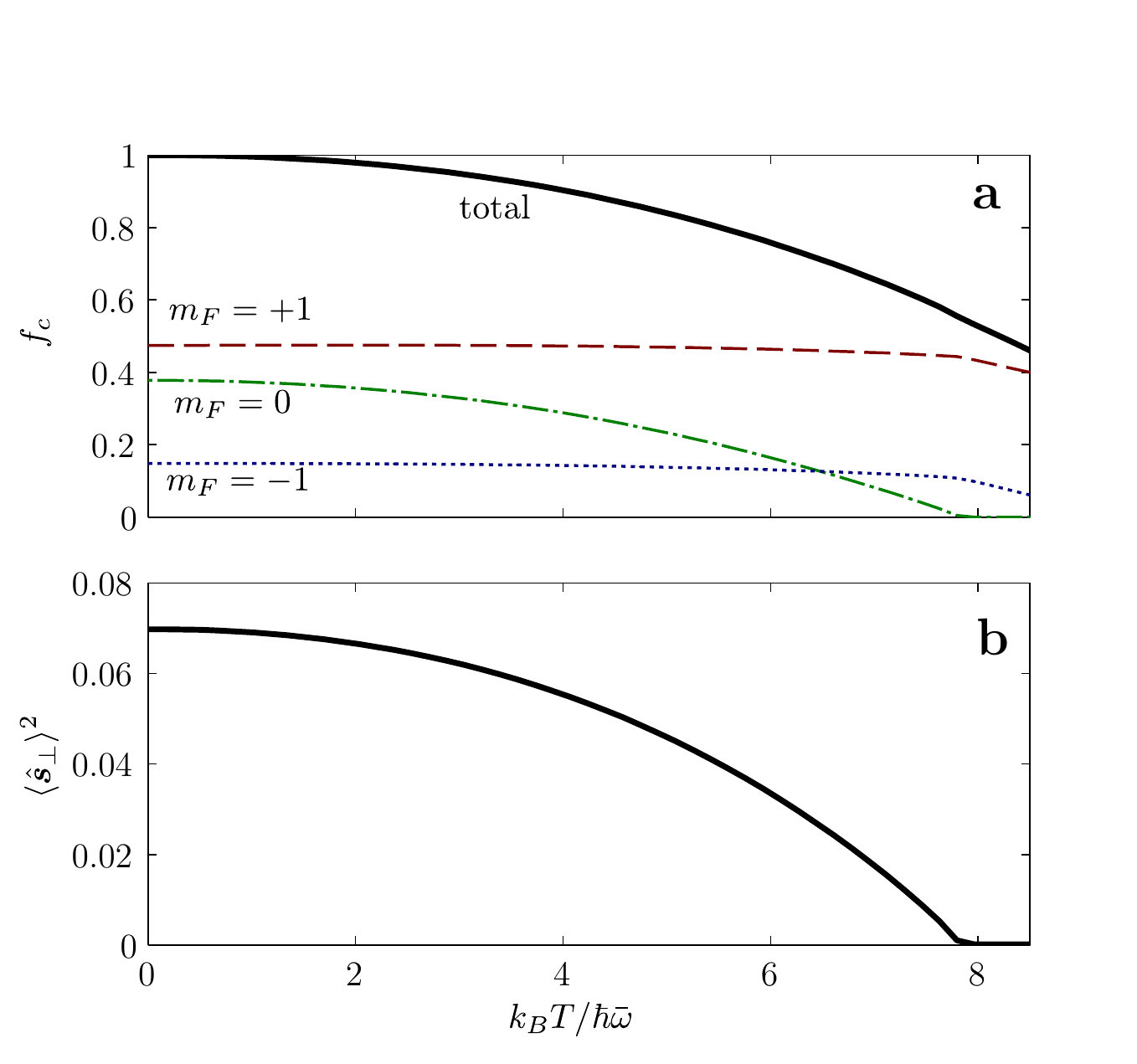}
\caption{(Color online) {\bf(a)} Partial condensed fractions $f_c^{(m_F)}$ for each Zeeman component $m_F=+1$ (dashed red), $m_F=0$ (dash doted green) and $m_F=-1$ (dotted blue line) and total condensed fraction (black solid). Here $f_c^{(m_F)}=N_c^{(m_F)}/N$ is normalized to the total number of atoms. {\bf(b)} Transverse spin length $\langle \hat{\bm{s}}_\perp\rangle^2$ versus temperature. The calculation was done for a spherical trap of frequency $\bar{\omega}/(2\pi )=405$\,Hz, $N=7500$ atoms, $\overline{m}_z=0.33$ and $q/h = 6$\,Hz. In physical units, $k_B T=8\hbar\bar{\omega}$ corresponds to $T\approx 150 \,$nK.}
\label{fig:fcHF}
\end{figure}

We now discuss the thermal depletion of the condensate population. The single-mode approximation only describes the lowest energy ``spatial  mode'' into which the atoms condense. Higher energy modes can be thermally populated, leading to a {\it condensed fraction} $f_c=N_c/N$ lower than one. Here $N$ and $N_c$ denote respectively the total number of atoms and of condensed atoms, irrespective of their internal state. To describe the thermal component of the non-condensed cloud, we have adapted the Hartree-Fock (HF) description proposed in \cite{kawaguchi2012b} in the uniform case to our experimental situation (see Appendix\,\ref{app:HF} for details). 

The results of this calculation are shown in Fig.\,\ref{fig:fcHF} for parameters relevant to our experimental situation, where we plot the partial condensed fractions for each Zeeman component $f_c^{(m_F)}$, defined as the ratio of condensed atom number in state $m_F$ to the total atom number. The condensed fraction in $m_F=0$ decreases first. Above  $k_B T\geq 7.8 \hbar \omega$, the $m_F=0$ component is purely normal and the condensate is formed by $m_F=\pm 1$ only. As found in \cite{kawaguchi2012b}, the contribution of the thermal component to the average spin vector is oriented opposite to the average spin of the condensate. The total transverse spin is thus naturally reduced with increasing temperature \footnote{Note that Eq.\,(\ref{eq:A}) applies only for pure states and cannot be used directly at finite temperatures}. In the regime we have investigated, the temperatures fulfill $k_B T \gg q, U_s$. As a result, the non-condensate spin vector is always much smaller in magnitude than its condensed counterpart, and we find that the main effect that reduces the length of the transverse spin vector is the reduction of the condensed fraction. The results of Section\,\ref{sec:equilibrium} can be directly used, provided one replaces the total atom number $N$ by the condensed atom number $N_c<N$ and the reduced populations $n_{m_F}$ by their condensed counterparts. For a total condensed fraction $f_c=0.8$, $\langle\hat{\bm{s}}_\perp\rangle^2$ is reduced to about $57\%$ of its zero temperature value. 



\section{Spin noise reveals spin-nematic order}
\label{sec:spinnoise}

In contrast to the phase $\Theta$, which is locked to $\pi$ in equilibrium by the spin-exchange interactions, the phase $\alpha$ is expected to take random values from one realization to the next. When dealing with many realizations of the same experiment, the initial many-body state is thus characterized by a statistical mixture  
\begin{align}
\label{eqrho}
\hat{\rho} = \int_0^{4\pi} \frac{d \alpha}{4\pi}\, \vert  \bm{\zeta}^N \rangle \langle   \bm{\zeta}^N\vert
\end{align}
rather than a pure state $\vert \bm{\zeta}^N \rangle $ with $N$ bosons in the spin state $\vert \bm{\zeta}\rangle$. Only three parameters (\textit{e.g.}, $n_0,m_z,\langle\hat{\bm{s}}_\perp\rangle^2$) are needed to characterize the ensemble, down from four to specify completely each member $\vert \bm{\zeta} \rangle$. In spite of the randomness of the spin orientation, these three parameters can still be measured using spin rotation provided one goes beyond single-particle observables and measures spin noise (recent experiments used similar techniques to reveal squeezing \cite{lucke2011twin,gross2011atomic,hamley2012spin,lucke2014detecting}). 

Figure\,\ref{Fig1}c-e illustrates the method geometrically in terms of the mean spin vector $\langle \hat{\bm{s}} \rangle$. 
The mean spin vector for a general spin 1 pure state $\vert \bm{\zeta}\rangle$ lies on or inside a sphere of radius one, with the phase $\alpha$ describing the azimuthal angle of the transverse component of the mean spin vector (panel c). 
The ensemble of possible initial states with a uniform distribution for $\alpha$ lie on a circle of radius $\vert \langle\hat{\bm{s}}_\perp \rangle\vert$ around the $z$ axis (panel d). In order to measure this radius, we rotate the state by a known angle $\Omega t$ around the $y$ axis and measure the magnetization $m_z'$ after rotation (panel e). As seen from the figure, the initial fluctuations of the transverse orientation map to fluctuations of $m_z'$, which are readily measured. 

For a more quantitative description, we use the standard angular momentum algebra to obtain the rotated operator
\begin{align}
\hat{S}_z' = \hat{R}_y^\dagger(\Omega t ) \hat{S}_z \hat{R}_y(\Omega t )= \cos(\Omega t ) \hat{S}_z - \sin(\Omega t ) \hat{S}_x.
\end{align}
Here and in the following, primed variables denote quantities evaluated after the spin rotation is complete. We now introduce a key assumption: the initial density matrix is invariant under rotation around the $z$ axis. This is satisfied in particular by the density matrix in Eq.\,(\ref{eqrho}), with a random phase $\alpha$ uniformly distributed in $[0,4\pi]$. The value of an observable measured after averaging over many realizations of the experiment is
\begin{align}
\langle \hat{O}  \rangle_\alpha & = \frac{1}{4\pi}\int_0^{4\pi} d\alpha~\langle \mathcal{R}^\dagger\hat{O} \mathcal{R} \rangle,
\end{align} 
where $\mathcal{R}=e^{-i\Omega t \hat{S}_y} e^{-i\frac{\alpha}{2} \hat{S}_z}$. The $\langle\;\cdot\;\rangle_\alpha $ symbol stands for a double average : the first one, denoted by $\langle  \;\cdot\; \rangle$, is the usual average over the quantum state before rotation for each realization, and the second one is done over random values of $\alpha$ arising from one experimental realization to the next. Defining an average in this way allows us to obtain formula expressing measurement results without specifying the initial state. 

Using this result, we find the average magnetization after the pulse,
\begin{align}
\label{eq:Meanmz}\langle m_z'\rangle_\alpha  & =\frac{1}{N} \langle \hat{S'}_z  \rangle_\alpha = \cos(\Omega t ) m_z,
\end{align}
which is independent of $\langle\bm{s}_\perp \rangle$. However, the variance of the same quantity is given by
\begin{align}
\Delta m_z'^2 &=  \cos^2(\Omega t ) \Delta m_z^2 + \frac{1}{2N^2} \sin^2(\Omega t ) \langle \hat{S}_x^2+\hat{S}_y^2 \rangle,
\end{align}
where
\begin{align}\nonumber
 \hat{S}^2_x+\hat{S}^2_y =& N + \hat{N}_0 + 2\hat{N}_0 \left( \hat{N}_{+1}+\hat{N}_{-1}\right) \\ &+ \left( \hat{a}_{+1}^\dagger\hat{a}_{+1}^\dagger \hat{a}_0^2 + {\rm h.c.} \right). \label{eq:Sx2Sy2}
\end{align}
In other words, relying only on the randomness of $\alpha$ we find that the variance of the magnetization $\Delta m_z'^2$ after the pulse measures the initial transverse spin fluctuations. This result holds for a short enough pulse, such that one can neglect any other terms than the oscillating field in the Hamiltonian during the evolution time. 

It is convenient to rewrite the variance as
\begin{align}
\nonumber 
\Delta m_z'^2 &=  \frac{1}{2} \sin^2(\Omega t ) \langle\hat{\bm{s}}_\perp\rangle^2\\
\label{eq:Stdmz}
& + \cos^2(\Omega t ) \Delta m_z^2 + \frac{1}{2} \sin^2(\Omega t ) \Delta s_\perp^2,
\end{align}
with $\langle\hat{\bm{s}}_\perp\rangle^2$ the squared length of the mean transverse spin, and with $\Delta s_\perp^2=\langle \hat{S}_x^2+\hat{S}_y^2 \rangle/N^2-\langle\hat{\bm{s}}_\perp\rangle^2$  its variance. 
For a spinor condensate with $\langle\hat{\bm{s}}_\perp\rangle \neq 0$, the term on the first line dominates over the smaller noise terms, and $\Delta m_z'^2 \approx  \frac{1}{2} \sin^2(\Omega t ) \langle\hat{\bm{s}}_\perp\rangle^2$. We thus expect that the variance $\Delta m_z'^2$ oscillates with the rotation angle $\Omega t $ and reaches its maximum for $\Omega t  =\pi/2$ where the slope of $m_z'$ versus $\Omega t $ is maximum. In our experiment, the last two noise terms in Eq.\,(\ref{eq:Stdmz}) are typically dominated by the preparation noise on $m_z$ (which also introduces noise on $n_0$ in the equilibrium state, and thus on $\langle\hat{\bm{s}}_\perp\rangle$). 

\section{Experimental techniques}\label{sec:exp}

\subsection{Condensate preparation}

We prepare spinor condensates in a well-controlled homogeneous static magnetic field $\bm{B}$ oriented along the $z$ axis [see Fig.\,\ref{Fig1}a]. We start from a precooled thermal cloud of $^{\rm 23}$Na atoms in a crossed optical dipole trap \cite{jacob2011a}. The atomic cloud is partially magnetized, with a magnetization $\overline{m}_z \approx 0.5$ on average resulting from previous cooling steps. We adjust the magnetization by either demagnetizing the atoms further with near-resonant RF-magnetic field sweeps, or by magnetizing it by evaporation in a magnetic field gradient (``spin distillation'') \cite{jacob2012a}. We are able to produce final magnetizations ranging from $\overline{m}_z=0$ to $\overline{m}_z=1$, with a typical error of $2-3\,$\%. 

 After preparing a spin mixture well above the critical temperature for Bose-Einstein condensation, the depth of the optical trap is lowered in a few seconds to perform evaporative cooling. A hold time of 3\,s is added after the end of the ramp to ensure that the cloud reaches equilibrium \cite{jacob2012a}. At the end of the evaporation ramp, the atoms are confined in the crossings of the two beams of the dipole trap, where the trapping potential is well-approximated by a harmonic trap with average trap frequency $\overline{\omega}/2\pi \sim 405\,$Hz (the trap frequencies are in the ratio $1:0.85:0.5$). 

Experiments reported in this article are performed with ``almost pure'' Bose-Einstein condensates (BECs) containing typically 7500 atoms at a 
trap depth $V_T/k_B \approx 400\,$nK. By ``almost pure'', we mean that no discernible thermal component can be observed in absorption images. 
The measured condensed fraction $f_c=N_c/N$ is usually obtained by fitting a bimodal profile to absorption images \cite{ketterlereview}. In our experiment, the contribution of the thermal component becomes difficult to detect for condensed fractions larger than $f_c\approx 0.8$, and the bimodal fitting procedure unreliable. This sets a lower bound $f_c \geq 0.8$ on the condensed fraction for the experiments presented in this article.  


We probe the sample using absorption imaging after free expansion in a magnetic field gradient, as shown in Fig.\,\ref{Fig1}b, and measure the normalized populations $n_{m_F}$ of each Zeeman component $m_F=0,\pm 1$ \cite{stamperkurn2013a}. The three Zeeman components are imaged after releasing the cloud from the trap in the presence of a magnetic force separating the Zeeman components. Specifically, we apply a quadrupole field $\bm{B}_{\rm q}=b'(2x {\bf e}_x-y{\bf e}_y-z{\bf e}_z)$ together with a uniform ``separation'' field $B_x {\bf e}_x$, with $b'\approx 7\,$G/cm and $B_x\approx3\,$G. The resulting adiabatic magnetic potential is given by $U_{\rm mag}=g_F m_F \mu_B \vert B_x {\bf e}_x +\bm{B}_{\rm q} \vert \approx g_F m_F \mu_B \vert B_x \vert +g_F m_F \mu_B b' x +\cdots$, with $g_F=-1/2$ the Land\'e factor and with $\mu_B$ the Bohr magneton. The quadrupole and separation field are ramped up in a few milliseconds, while the bias field $B \bm{e}_z$ applied during the experiment is simultaneously ramped down. 

\subsection{Experimental implementation of Rabi oscillations}
\label{sec:Rabi}
\begin{figure*}[ht!!!]
	\begin{tabular}{lcr}
	 	\includegraphics[width=0.5\textwidth]{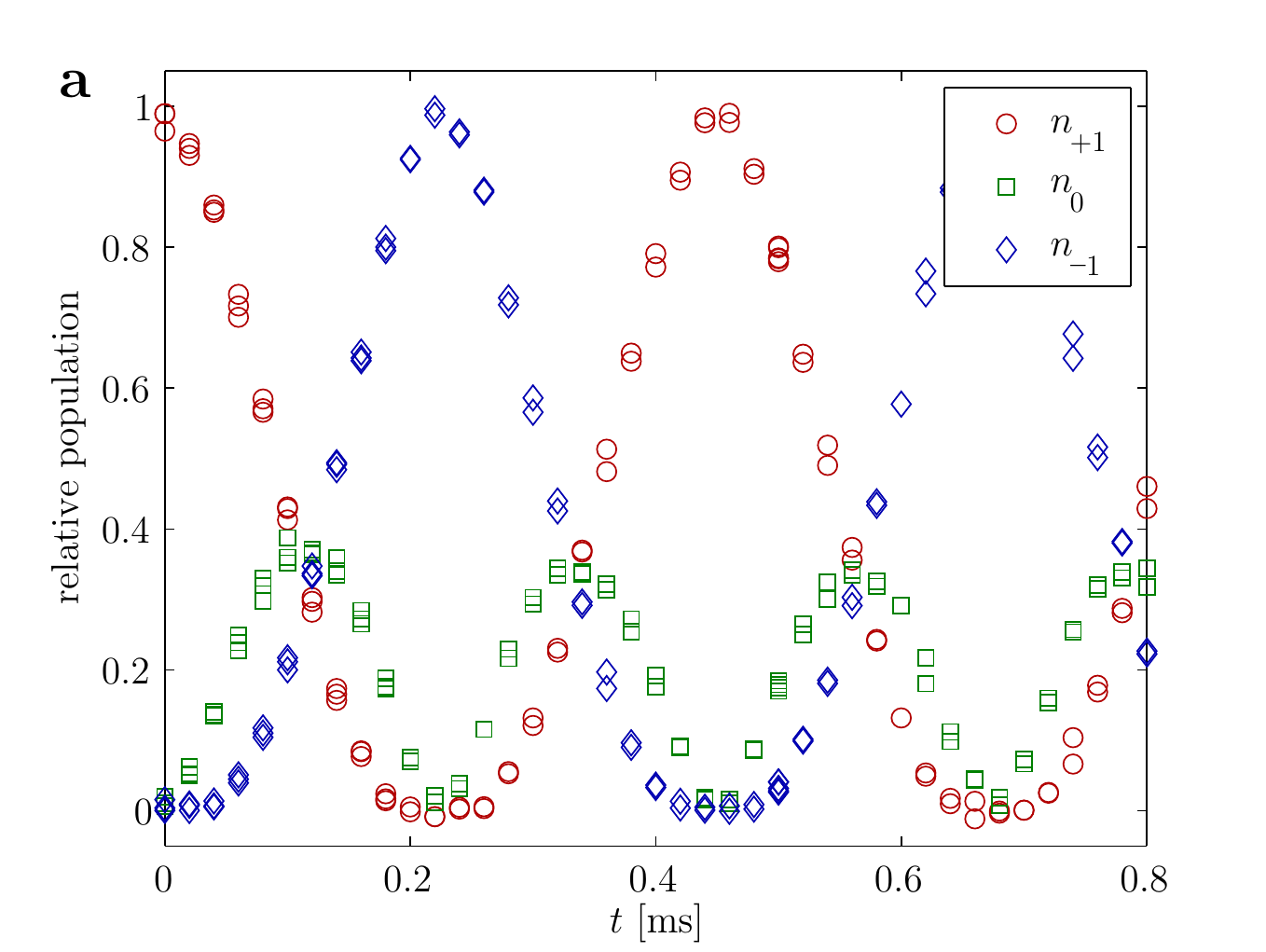}&
		\includegraphics[width=0.5\textwidth]{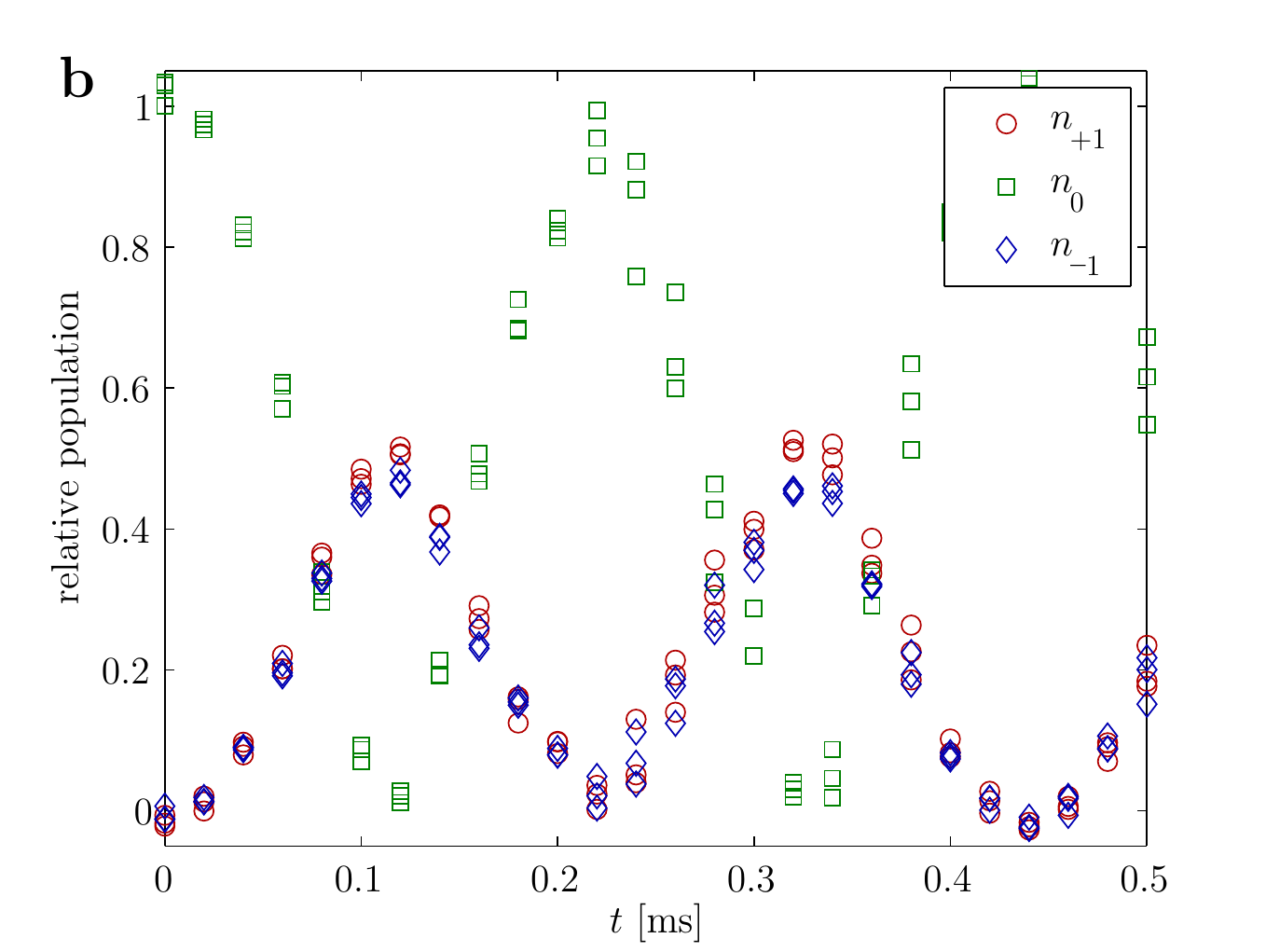}
		\end{tabular}    
	\caption{(Color online) Rabi oscillation starting either from a state with all atoms in $m_F=+1$ {\bf(a)} or $m_F=0$ {\bf(b)}. The residual fluctuations are dominated by preparation noise, imperfections in the Rabi rotation parameters and the detection noise, all with roughly comparable contributions.}
	\label{fig:RabiOsc}
\end{figure*}

We apply a spin rotation using a radio-frequency (RF) magnetic field along $\bm{y}$ oscillating at the Larmor frequency. This RF field induces Rabi oscillations with Rabi frequency $\Omega$. After a certain evolution time $t$ which determines the rotation angle $\Omega t$, we measure the final populations $n_{m_F}^\prime$ after spin rotation. The bias field is small enough to neglect the quadratic Zeeman shift ($q<100$\,Hz) compared to the Rabi frequency ($\Omega/2\pi\sim 5$\,kHz). At the end of the pulse, the separation field $B_x \bm{e}_x$ is increased first, folllowed by the magnetic gradient used for SG imaging and by the decrease of the bias field $B_z \bm{e}_z$. The timing of the sequence is shown in Fig.\,\ref{fig:MFdynamics}a. Ramping up the separation field $B_x$ is done with a linear ramp of $T=3$ ms duration, sufficiently slowly to remain adiabatic with respect to spin flips ($\omega_L T \ll 1$). The optical trap is switched off 10 ms after the end of the RF pulse (see Section \ref{sec:MFdynamics} below). 

We have tested this sequence in two special cases, where all the atoms are initially in the $m_{F}=+1$ state and or in the $m_{F}=0$ state. We are able to prepare these two states with little preparation noise, $\Delta m_{z}\lesssim 1\%$. 
The measured oscillations are presented in Fig.\,\ref{fig:RabiOsc}. The contrast is close to $100\,$\%, and we do not observe any sizeable dephasing of the oscillations after several Rabi periods. This shows that the assumption of adiabatic following when ramping up the different magnetic fields is valid. 

\subsection{Influence of spin mixing after the spin rotation}
\label{sec:MFdynamics}

\begin{figure}[htb]
	\centering	 	
\includegraphics[width=0.5\textwidth]{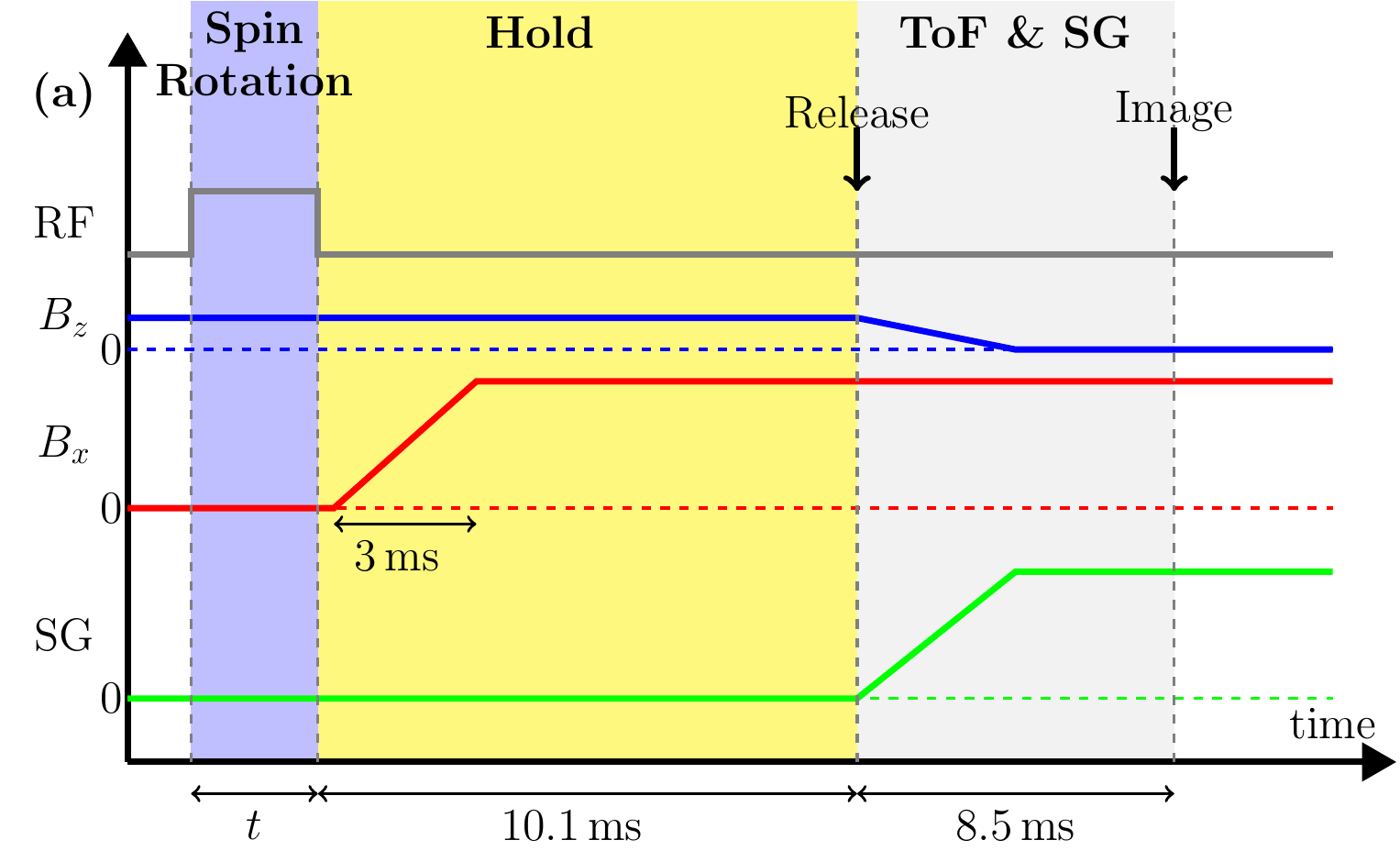}\\
	 	\includegraphics[width=0.5\textwidth]{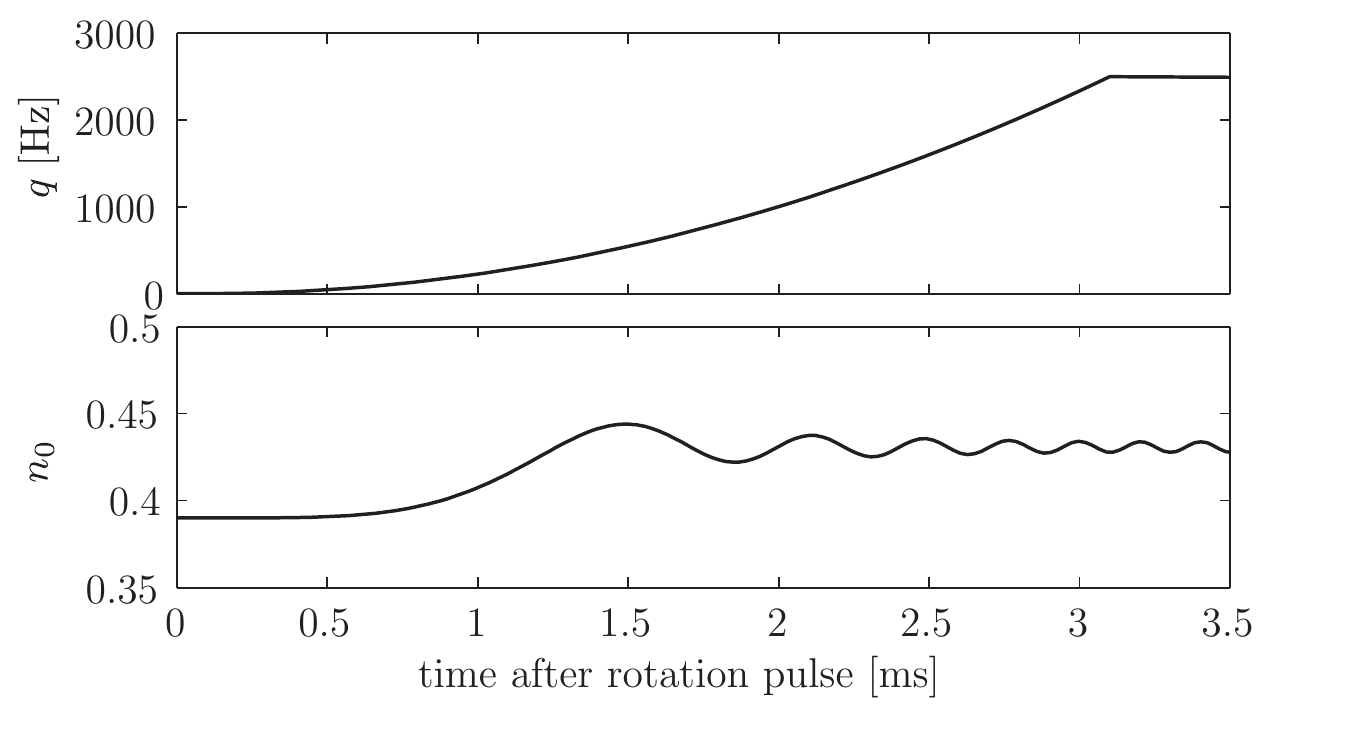}
	\caption{(Color online) {\bf (a)}: Schematic diagram (not to scale) showing the experimental sequence. ``RF'' indicates the rf pulse inducing spin rotations, $B_z$ is the bias field applied before and during the spin rotation, $B_x$ and ``SG'' denote respectively the ``separation field'' and magnetic field gradient required for SG imaging. {\bf (b)} The ramp of $B_x$ after the spin rotation results in a time-dependent Quadratic Zeeman energy (QZE) $q$ increasing within $3\,$ms after the end of the rf pulse (top panel). The evolution after the spin rotation of the normalized population $n_0$ due to spin-mixing interactions calculated from Eq.\,(\ref{eq:spinmixing}) is shown in the lower panel for an initial phase $\Theta_i=\pi$, and an initial population chosen such that the final population is  $n_0\approx0.43$ (as measured for data set \textit{a}), and $U_s/h= 38$~Hz.}
	\label{fig:MFdynamics}
\end{figure}  

The sudden change of the spin state due to the spin rotation should in principle trigger a spin oscillation dynamics \cite{pu1999a,chang2005a,kronjaeger2005a,zhang2005a,black2007a} driven by spin-exchange interactions during the 10\,ms hold time following the spin rotation. As seen before, the applied magnetic field is also changed after the spin rotation, from $\bm{B}=B_z\bm{e}_z$ to $\bm{B}=B_z\bm{e}_z+B_x\bm{e}_x$. The quadratic Zeeman energy $q$ increases during this ramp, according to the curve shown in Fig.\,\ref{fig:MFdynamics}b. This increase is fast compared to the time scale set by spin-exchange interactions, $h/U_s\sim 25\,$ms, and it reduces spin-mixing dynamics due to exchange collisions that would otherwise develop during the 10\,ms hold time after the RF pulse. 

Nevertheless, a residual dynamics still takes place and modifies slightly the population $n_0$ measured in SG imaging. Note that the effect of the spin interaction during the RF pulse is negligible ($U_s / \hbar\Omega \sim 0.008 $).
We model the spin-mixing oscillations using the theoretical framework given in \cite{zhang2005a} (see Appendix\,\ref{app:spinmixing}). An example for $\Theta_i=\pi$ is shown in Fig.~\ref{fig:MFdynamics}b. The main changes in $n_0$ occur early in the ramp. Once $q$ has settled at its final value $q_f\sim h \times 2.5\,$kHz, the dynamics continue as a small amplitude oscillation of the population $n_0$ around an offset value (the so-called quadratic Zeeman regime \cite{kronjaeger2005a}). The oscillation amplitude is small ($\sim U_s/ q_f \sim 0.015$) and comparable to our detection noise. Changing the magnetic field to higher values would further reduce the amplitude without significantly changing the offset of $n_0$. Taking the long-time offset as the measured value of $n_0$, we find that the effect of the ramp amounts to {\it increase} the relative population in $n_0$ from its initial value by up to 0.05 for an initial angle $\Theta_i=\pi$, a small but measurable change.  

We emphasize that the spin-mixing dynamics does not change the magnetization $m_z$ of the system, but only the individual populations $n_{m_F}$. Therefore, the occurrence of spin mixing does not influence the analysis of the variance of $m_z$ after spin rotation in Section\,\ref{sec:spinnoisemeasurement}. On the other hand, it does affect the maximum likelihood analysis, as detailed further in section \ref{sec:MLE}.

\subsection{ Determination of $U_s$ from spin-mixing dynamics }
\label{sec:Us}
\begin{figure}[ht!!!]
	\begin{tabular}{lcr}
	 	\includegraphics[width=0.5\textwidth]{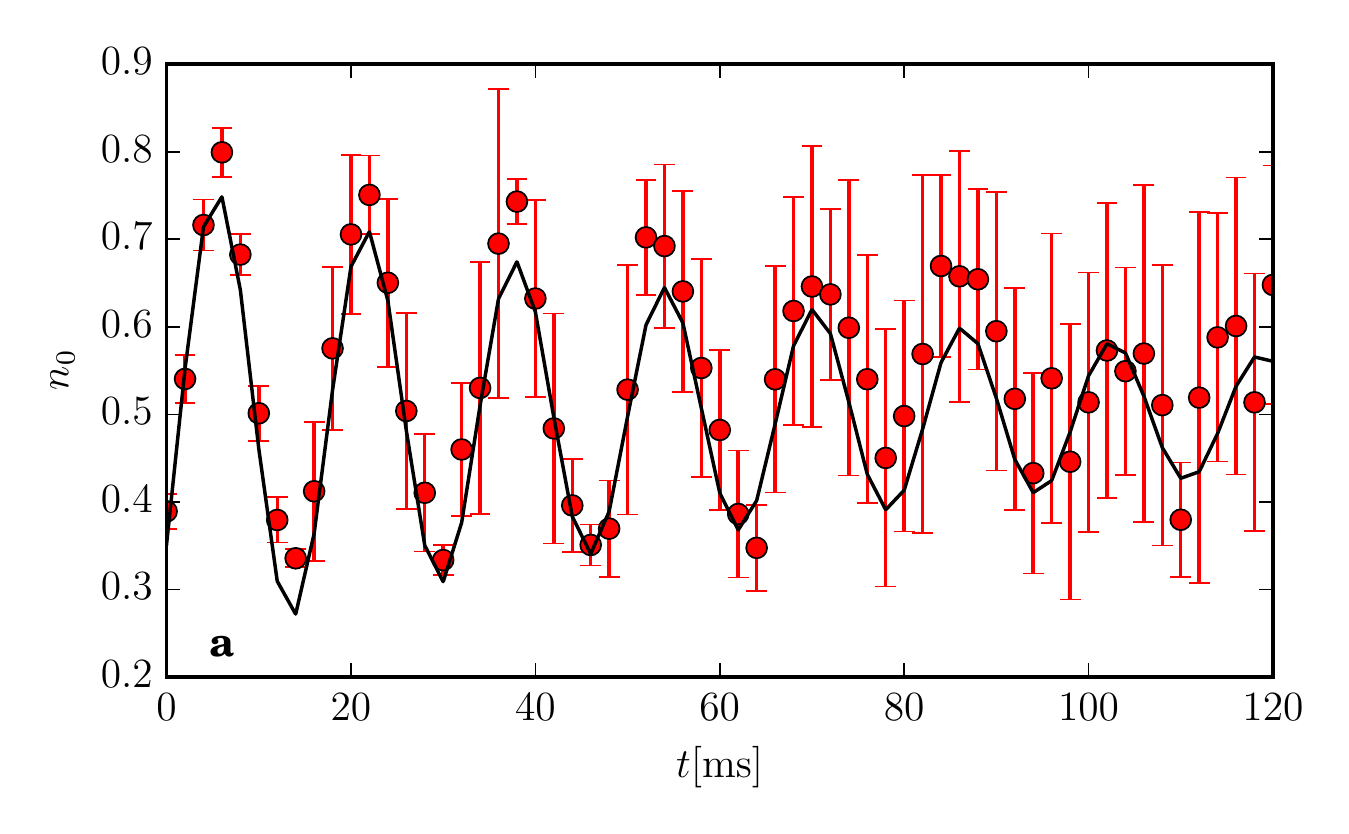}\\
		\includegraphics[width=0.5\textwidth]{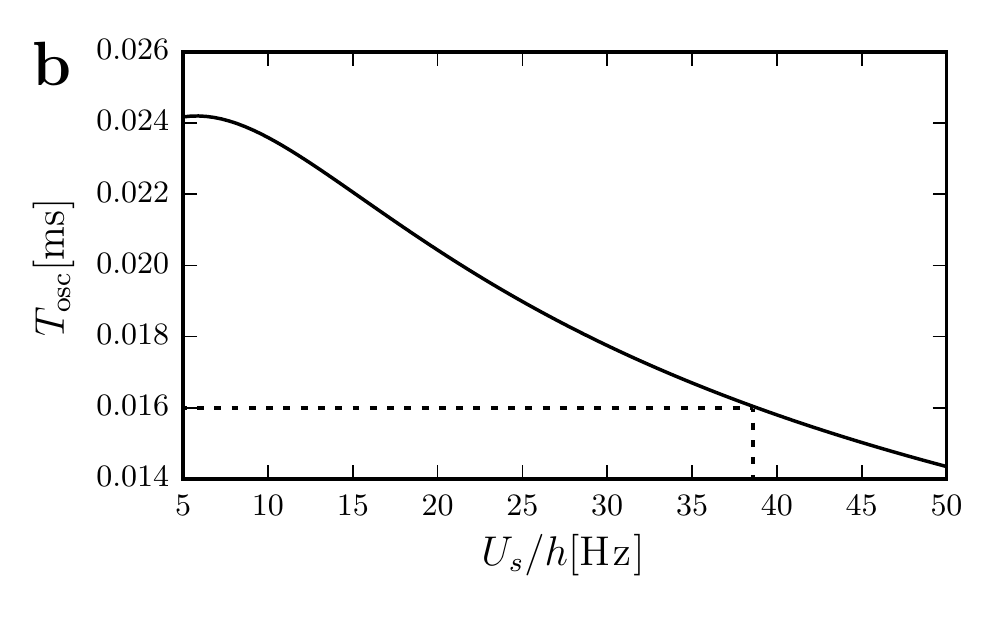}
		\end{tabular}    
	\caption{(Color online) Spin-mixing oscillations {\bf(a)} and calculated oscillation period {\bf(b)}. A fit to a damped sinusoid is shown in {\bf (a)} as solid line, and yields an oscillation period $T_{\rm osc} \approx 16\,$ms indicated by dashed lines in {\bf (b)}.}
	\label{fig:SpinOsc}
\end{figure}

We have measured directly the exchange interaction parameter $U_s$ by deliberately inducing spin-mixing dynamics and recording the oscillations of the normalized population $n_0$ after a sudden change (see Fig.\;\ref{fig:SpinOsc}a). Starting from a condensate with all atoms in the $m_F=0$ state, prepared as explained above at a bias field $B\approx 282 \,$mG [$q/h \approx 22\,$Hz], we first apply a spin rotation to produce a mixture with roughly balanced populations in all Zeeman states. This results in an initial state as given by Eq.\,(\ref{eq:Psi}), with $n_{0}\approx 0.38$ and $m_z\approx 0$. Spin-changing collisions produce high-contrast oscillations in the Zeeman populations, as observed in previous work for $m_z\neq 0$ \cite{chang2005a,kronjaeger2005a,black2007a,liu2009a}. The oscillation period has been predicted analytically in \cite{zhang2005a}, and is a function of $ n_{0},m_z,q$, which are known, and of $U_s$, which is not. We extract $U_s/h \approx 38 \,$Hz from the measured period $T_{\rm osc}\approx 16\,$ms (see Fig.\,\ref{fig:SpinOsc}b). 

\section{Spin noise measurement of spin-nematic order}
\label{sec:spinnoisemeasurement}

\begin{figure*}[t]
  
  \centering
    \includegraphics[width=\textwidth]{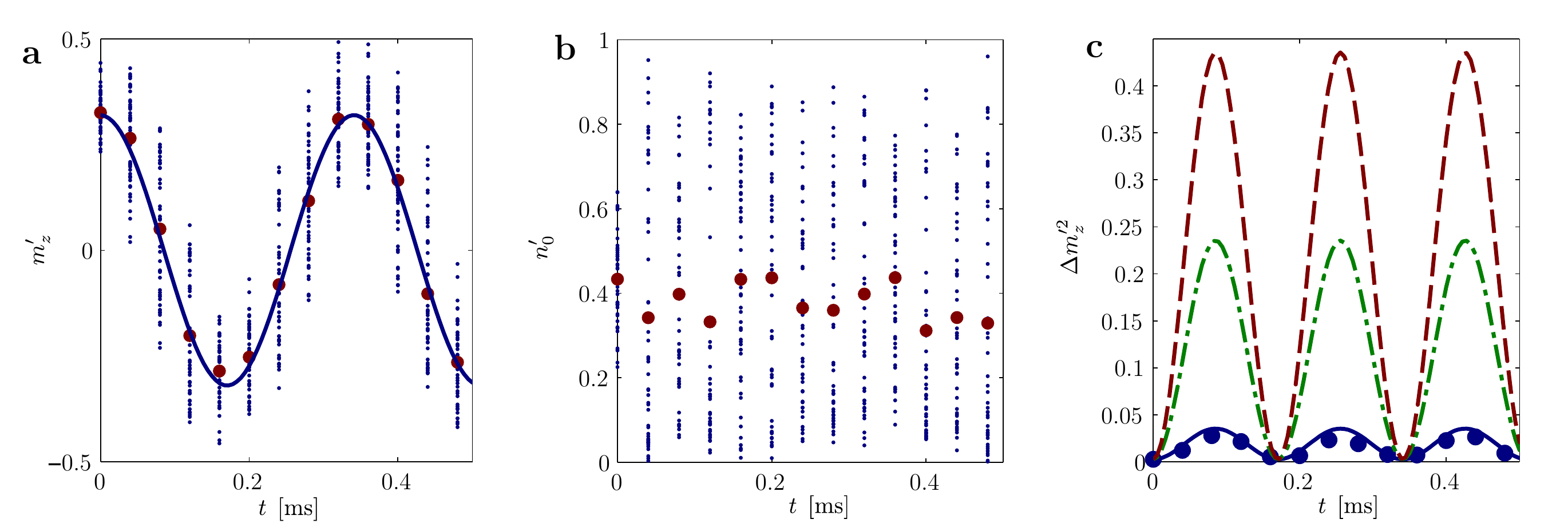}
	   \caption{(Color online) {\bf (a)}: Magnetization $m'_z$ and {\bf (b)}: relative population $n'_0$ in the $m_F=0$ state versus duration of the Rabi pulse (or equivalently, rotation angle). The small blue dots correspond to single-shot measurements, while the larger red circles correspond to the average $(\overline{m}'_z,\overline{n}'_0)$ over all measurements for each pulse duration. The solid line shows a cosine fit to the average $\overline{m}'_z$, from which we extract the Rabi frequency $\Omega$. {\bf (c)}: Variance of $m'_z$ (blue circles) versus duration of the Rabi pulse oscillating at twice the Rabi frequency. The blue solid line corresponds to the theoretical prediction at zero temperature obtained from Eqs.~(\ref{eq:Sperp},\ref{eq:Stdmz}) and an initial phase $\Theta=\pi$. The red dashed line ($\Theta=0$) and green dash-dotted line (random $\Theta$) are shown for illustrative purposes. The data are from set \textit{a}, with $\overline{m}_z=0.33$ and $q/h=6.0$ Hz ($B=147\,$mG).}
    \label{Fig2}
\end{figure*}

We now describe our experimental results on the measurement of the transverse spin using spin noise, as described in Section\,\ref{sec:spinnoise}. In total we have taken three different data sets for different initial magnetizations and magnetic fields which we label $\textit{a},\textit{b},\textit{c}$ (see Fig.\,\ref{Fig5}). The first two cases are above the $T=0$ phase transition, while the third one is below. In each case, we drive Rabi oscillations with Rabi frequency $\Omega$ for an evolution time $t$, as described for quasi-pure spin states in Section\,\ref{sec:Rabi}, and record the evolution of the relative populations $n_{m_F}^\prime$ after spin rotation. 

\subsection{Magnetization variance above the phase transition}

We first focus on data set \textit{a}. Fig.\,\ref{Fig2} shows typical raw data for the relative magnetization $m_z'$ (a) and the relative population $n_0'$ (b) for different rotation times $t$. As a result of the random orientation of the transverse spin (due to the random nature of $\alpha$), large shot-to-shot fluctuations of the individual populations are observed. The mean magnetization behaves as predicted in Eq.\,(\ref{eq:Meanmz}). We extract the Rabi frequency $\Omega$ from a cosine fit to the mean population $m'_z$ (see Fig.\,\ref{Fig2}a). 

Fig.\,\ref{Fig2}c shows the variance of $m'_z$, displaying the expected oscillations at twice the Larmor frequency. We compare the experimental results to the prediction of Eq.~(\ref{eq:Sperp},\ref{eq:Stdmz}) (blue solid line). The transverse spin length $\langle\hat{\bm{s}}_\perp\rangle^2$ is computed with $\Theta=\pi$, with the measured $\overline{m}_z$ and with the population $n_0^\ast$ found by minimizing $E_{\rm MF}$ \footnote{For this comparison we use Eq.~(\ref{eq:Stdmz}). Noise in $m_z$ was deduced from the measured distribution in the initial state. Noise in $\langle\hat{\bm{s}}_\perp\rangle$ was deduced from this measurement and Eq.~(\ref{eq:Sperp}) for $\Theta=\pi$.}. For comparison, we also show the transverse spin length for the same $\overline{m}_z, n_0^\ast$ but $\Theta=0$ (red dotted line) and for random $\Theta$ with uniform distribution (green dash-dotted line), that would correspond to a ferromagnetic system and to a non-interacting system (no phase locking), respectively. Our measurements are best described by $\Theta=\pi$, as expected for antiferromagnetic systems in equilibrium. This shows that the system attempts to minimize its transverse spin, or equivalently maximize its alignment, thereby revealing spin nematic ordering. 

As seen from Fig.\,\ref{fig:VarData}b,c, data sets \textit{b}, \textit{c} show the same behavior as the case \textit{a} discussed above, an oscillation of the variance with fixed amplitude. Data \textit{b} is qualitatively comparable to \textit{a}. Case \textit{c}, taken below the $T=0$ phase transition, deserves a separate discussion which we defer to Section\,\ref{sec:variancec}.

  %

\subsection{Spin thermometry}

We attribute the slight difference between the measured amplitudes of the variance oscillations and the prediction of Eq.~(\ref{eq:Stdmz}) for $\Theta=\pi$ in Fig.\,\ref{Fig2}c to a non-zero temperature. We addressed this point in details for data set \textit{a} using the Hartree-Fock treatment of Section\,\ref{sec:finiteT}. Generally, we have found that increasing the temperature reduces the transverse spin per atom. Experimentally, the condensed fraction can only be estimated as $f_c \gtrsim 0.8$ (see Section\,\ref{sec:exp}). We show in Fig.\,\ref{fig:VarData}a a shaded area where the lower limit corresponds to $f_c=0.8$ and the upper one to $f_c=1$, indicating that even a small non-condensed fraction leads to a measurable decrease of the oscillation amplitude. In fact, the oscillation variance can be seen as a low-temperature thermometer. A temperature $T \approx 80\,$nK (condensed fraction $f_c\approx 0.9$) is found to reproduce the observed oscillation level (dashed line in Fig.\,\ref{fig:VarData}a).


\subsection{Magnetization variance below the phase transition}

\label{sec:variancec}
For data set  \textit{c}, one would expect $\overline{n}_0=0$ and $\langle\hat{\bm{s}}_\perp\rangle^2=0$ according to the $T=0$ mean field picture. In contrast, we find a small initial population $n_0\approx 0.04$, and an oscillation of the magnetization variance with a small, but non-zero amplitude. The dotted lines in the figure correspond to the theoretical predictions which take the initial measured $n_0$ into account (corrected for the small shift in $n_0$ due to the spin changing collisions discussed in Section\,\ref{sec:MFdynamics}) and $\Theta=\pi$. 

A first explanation for this behavior could be the presence of the thermal (uncondensed) component. In a spinor BEC \cite{ueda2001a}, spin excitations are phase-locked to the condensed components, and a finite transverse spin originating from the uncondensed component could contribute to our signal. However, from the Hartree-Fock calculations described in Section\,\ref{sec:finiteT}, we found that the transverse spin of the uncondensed component remains very small for our typical parameters, and cannot explain the measured signal.

\begin{figure}
\centering
\includegraphics[width=\linewidth]{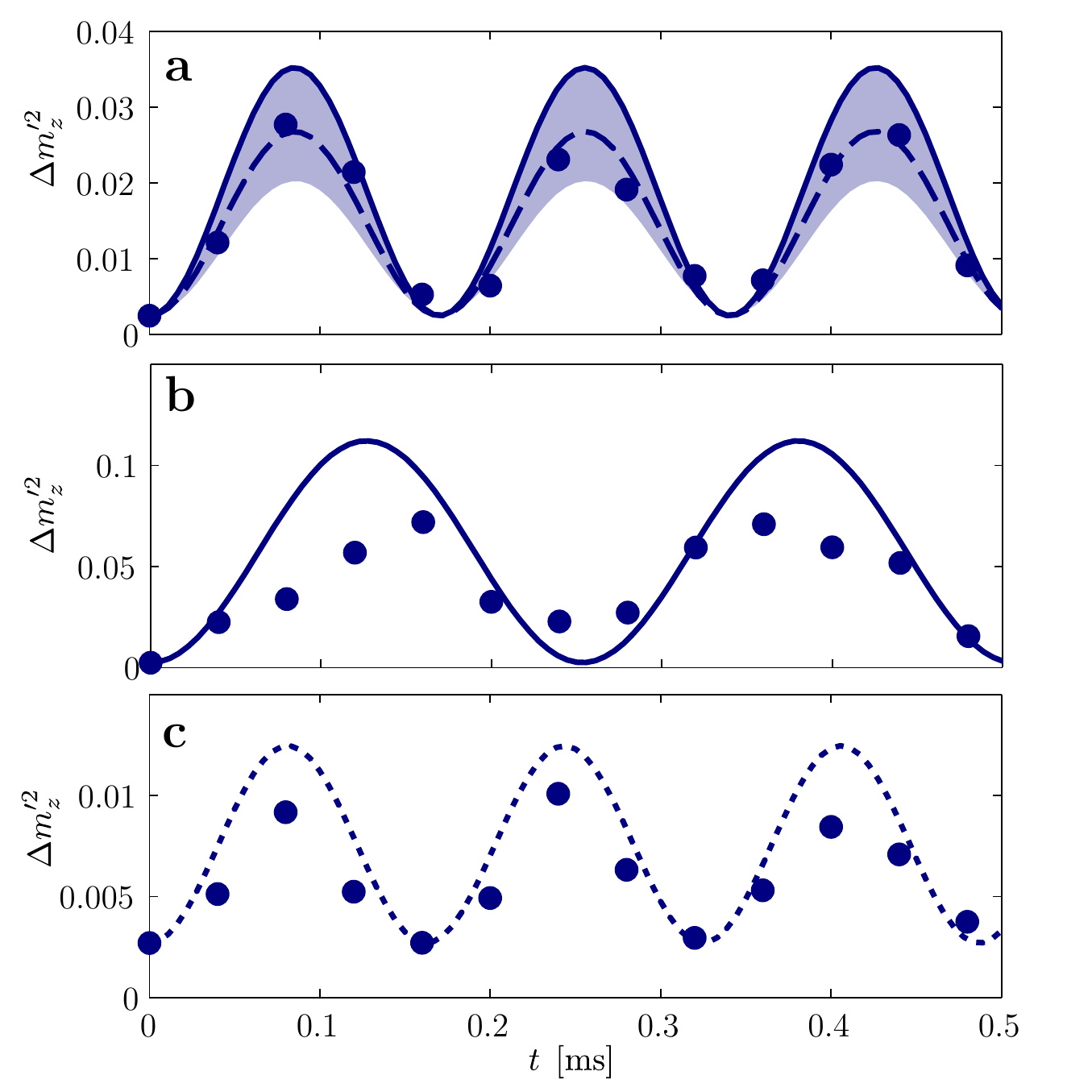}

\caption{(Color online) Close-up view of the magnetization variance for data sets \textit{a}, \textit{b} and \textit{c}. In {\bf{a}} and {\bf{b}}, the solid blue line is the zero temperature theory for an initial angle of $\Theta=\pi$ (antiferromagnetic interactions). In {\bf{a}}, the shaded area corresponds to the prediction of our Hartree-Fock model at finite temperature assuming a condensed fraction of $f_c\geq 80\%$. The dashed line gives the best agreement corresponding to a temperature $T \approx 80\,$nK. In {\bf{c}}, the data are shown for parameters below the $T=0$ phase transition line. The dotted curve is the theoretical expectation from Eq.\,(\ref{eq:Stdmz}) taking the initial population $n_0$ into account.}	
\label{fig:VarData}
\end{figure}

A second explanation comes from a finite temperature of the initial spin state of the condensate, which is then described by a statistical ensemble rather than a pure state as described in \cite{corre2015a} and Section\,\ref{sec:finiteT}. This leads to a finite population in $m_F=0$ even below the phase transition. By numerically integrating the thermal distribution described by the free energy given in \cite{corre2015a} for a typical temperature $T= 80$\,nK, we find a finite population $\overline{n}_0=0.016$. This leads to a maximal variance after rotation of $\Delta m_z'^2=0.005$, comparable to the oscillation amplitude of the variance in Figure\,\ref{fig:VarData}c. 

  %

\section{Maximum Likelihood estimation of the distribution of $\Theta$}
\label{sec:MLE}

\subsection{Principle of the method}

We now turn to a more general statistical analysis based on maximum likelihood estimation (MLE), which allows us to estimate the distribution of the angle $\Theta$ in a more quantitative way. It takes all available data into account, including the population $n_0'$ which was not used in the previous analysis. Given a set of measurements, the MLE method finds the most likely distribution among a set of parameter-dependent model distributions, thereby providing a statistical estimator for said parameters. 

We model the initial state by a density matrix
\begin{align}
\label{eqrho1}
\hat{\rho} = \int d \bm{\zeta} \, \frac{\mathcal{G}(n_0,m_z)P(\Theta)}{4\pi} \, \vert \bm{\zeta}^N \rangle\langle   \bm{\zeta}^N \vert,
\end{align}
with an integration measure $d \bm{\zeta}=dn_0 dm_z d\Theta d\alpha$. We assume for simplicity that the probability density functions $\mathcal{G}(n_0,m_z)$ and $P(\Theta)$ are Gaussians. We note that the equilibrium density matrix of a finite-temperature spin ensemble is well-approximated by Eq.\,(\ref{eqrho1}) with a Gaussian weight function \cite{corre2015a}. The joint probability density $\mathcal{G}(n_0,m_z)$ is peaked around the average value $(n_0^\ast,\overline{m}_z)$ with $n_0^\ast$ the population minimizing $E_{\rm MF}$, with a finite width mostly due to experimental imperfections in the preparation sequence. The covariance matrix characterizing $\mathcal{G}(n_0,m_z)$ is extracted from the experimental data. At $T=0$, $P(\Theta)$ is given by a Dirac delta, $P(\Theta) \propto \delta(\Theta-\pi)$, but acquires a finite width at finite $T$ (see Section\,\ref{sec:MLEdata} below). The mean value $\bar{\Theta}$ and standard deviation $\sigma_\Theta$ of $P(\Theta)$ are the unknown parameters to be estimated. Due to the periodic nature of $\Theta$, our choice is sensible only when $P(\Theta)$ is peaked around the mean, {\it i.e.} $\sigma_\Theta \ll 2\pi$. 

We use a Monte Carlo method to sample the initial distribution in Eq.\,(\ref{eqrho1}). For a given $\Theta$ and a measurement time $t_i$ (rotation angle $\Omega t_i$), the initial state $(n_0,m_z,\alpha,\Theta)$ is propagated in time using the rotation operator. Here we assume that the spin rotation is perfectly known, with rotation axis $y$ and a rotation angle extracted from the fit to $\langle m_z'\rangle_\alpha$ as before. Spin-mixing dynamics just after the spin rotation slightly change the relative population $n'_0$, and is taken into account in the propagation. After convolution of the final results with our known measurement noise, we get a conditional probability density $p_{t_i}(m'_z,n'_0 | \bar{\Theta},\sigma_\Theta)$ for the measured $(n_0',m_z')$. Given a set of independent observations $\{ m'_{z,i},n'_{0,i} \}$, we can construct a (log) likelihood function 
\begin{align}
\log L(\bar{\Theta},\sigma_\Theta)=\sum_{i} \log p_{t_i}(m'_{z,i},n'_{0,i}|\bar{\Theta},\sigma_\Theta). 
\end{align}
The distribution that accounts best for the observed results is found by maximizing this function. 

Since the estimator strongly depends on the chosen probabilistic model, it is important for this model to be close to the physical reality. In the following we motivate the model used in the MLE before discussing the results. 

\subsection{Model for the initial distribution}

\begin{figure}[htb]
\centering
\includegraphics[width=0.5\textwidth]{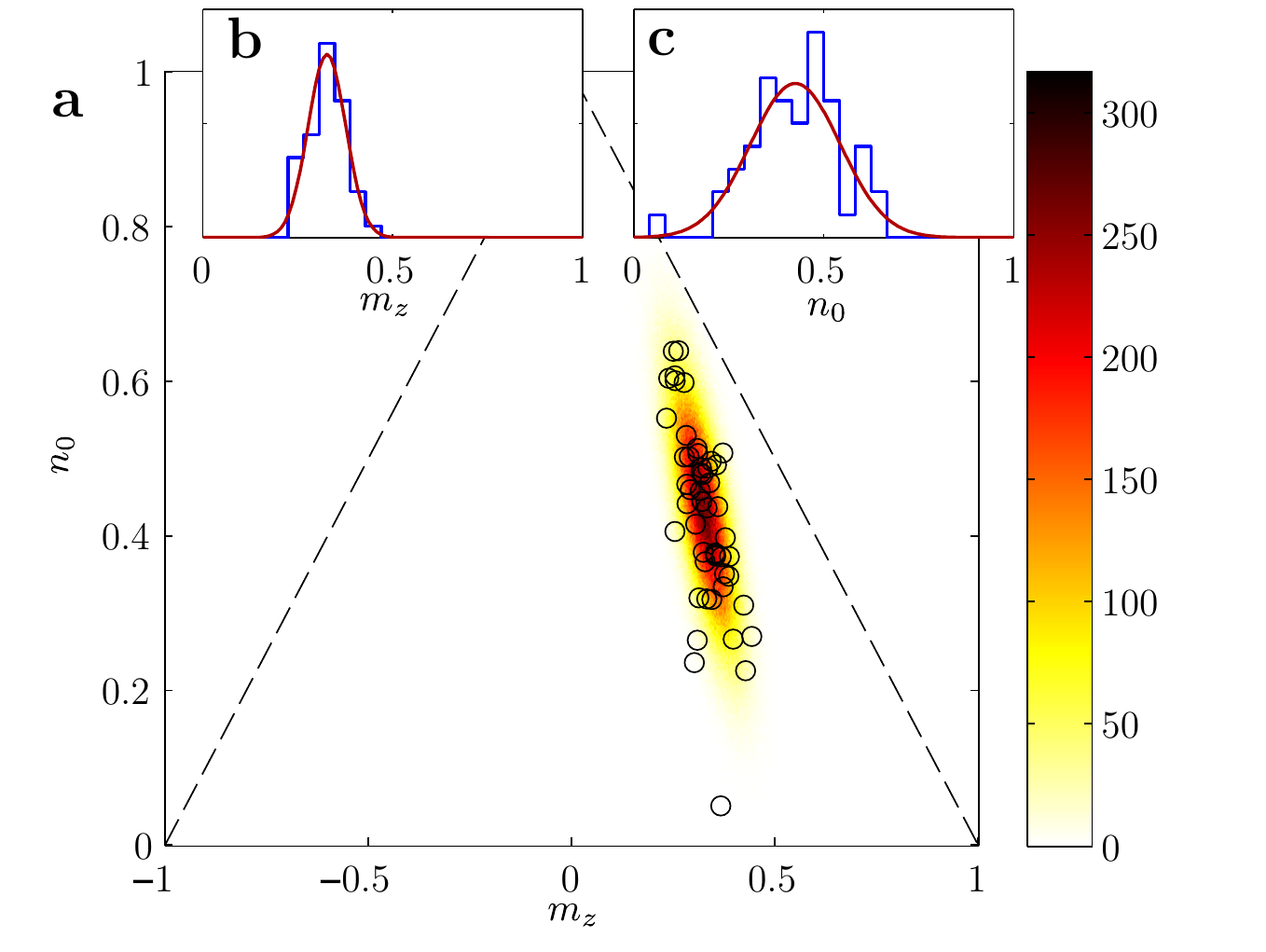}

\caption{(Color online) Initial distribution $\mathcal{G}(n_0,m_z)$ for data set \textit{a}. {\bf(a)} Measured initial populations are indicated by black circles. The color shows the two-dimensional histogram of the simulated $10^6$ initial points used in the Monte Carlo method. These points are drawn from a two-dimensional Gaussian distribution estimated from the initial measurements, and sampled on a square grid with a step size $2.5\times 10^{-3}$. The dashed lines limit the allowed area ($|m_z|\leq 1-n_0$). {\bf(b)}: Marginal histograms of measured initial populations for $m_z$, and {\bf(c)}: $n_0$. The red line indicates the marginal distributions used for the Monte Carlo analysis.}

\label{fig:InitialMLE}
\end{figure}

The distribution of initial states is probabilistic due to three different effects. The first effect is intrinsic to our theoretical model where the initial angle $\alpha$ takes random values from one realization to the next. The second probabilistic effect is due to experimental imperfections, mainly fluctuations of $m_z$ (from the preparation process and the subsequent evaporation), or fluctuations in the spin-spin interaction energy $U_s$ (due to fluctuations of the total atom number or of the confinement strength). Such fluctuations result in correlated fluctuations in $n_0$ due to the system exploring different minima of the mean field energy. We stress that the marginal distribution $P(\Theta)$ is {\it a priori} not affected by these fluctuations. A third random element originates from the finite spin temperature as described in Section\,\ref{sec:finiteT} which allows the system to explore states situated away from the minimum. The second and third effect are more pronounced close to the phase transition \cite{corre2015a}. 

We find empirically that the initial joint distribution of $n_0$ and $m_z$ in Eq.\;(\ref{eqrho1}) is well described by a two-dimensional Gaussian $\mathcal{G}(n_0,m_z)$. The mean and covariance matrix characterizing $\mathcal{G}$ are calculated from the measured data without spin rotation. We account for the spin changing collisions discussed in Section\,\ref{sec:MFdynamics}, which affect the measured ``initial distribution'', {\it i.e.} the distribution observed without any spin rotation. Specifically, for each values of $\Theta$, $m_z$ and $n_0$, the mean field equations (\ref{eq:spinmixing}) are used to find the initial value $n_{0,i}$ that leads to the measured one, $n_0(t=0)$. The known values of $q$ and the measured value of $U_s$ are used as fixed inputs for this calculation. The  initial distribution $\mathcal{G}(n_{0,i},m_z)$ deduced in this way is shown in Fig.\,\ref{fig:InitialMLE}. We estimate that experimental imperfections dominate the initial distribution $\mathcal{G}(n_0,m_z)$.

\subsection{Monte Carlo approach}
\label{sec:MLEMonteCarlo}

\begin{figure*}[htb]
\centering
\includegraphics[width=\linewidth]{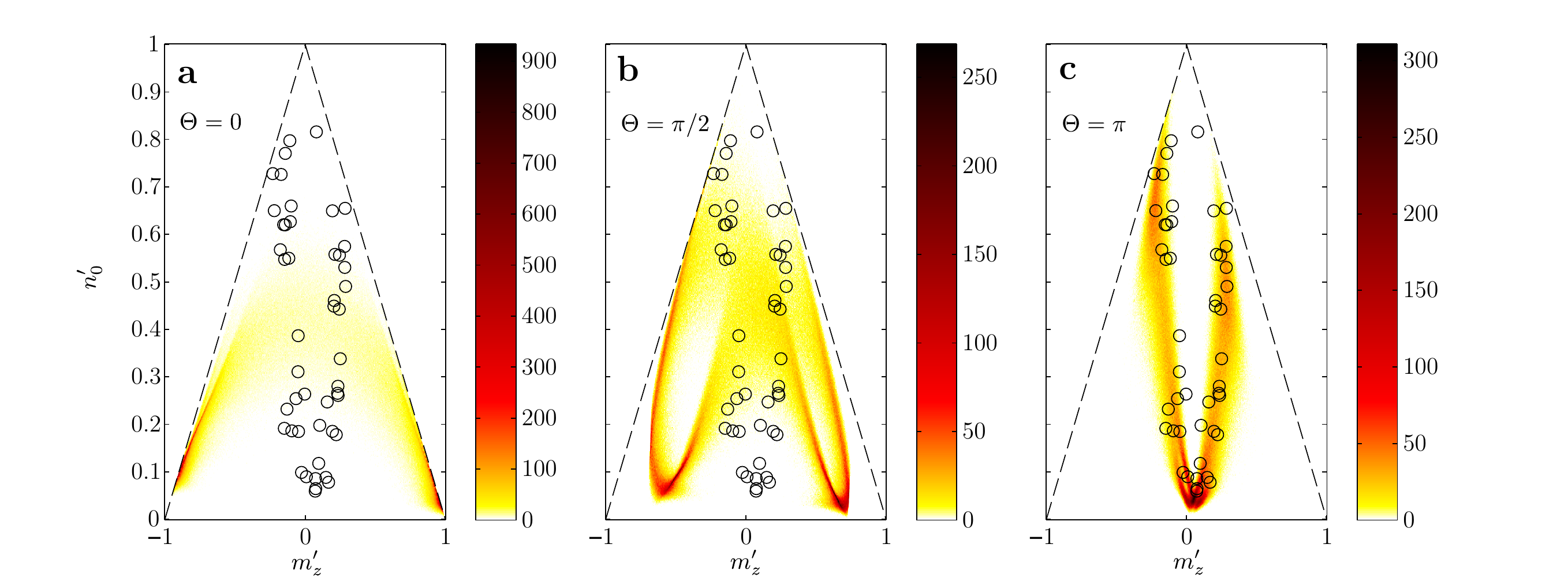}

\caption{(Color online) Comparison for data set $a$ of the Monte Carlo simulated populations with measured data (black circles) after Rabi-rotation for different assumed initial angles $\Theta$. The three panels show the color-coded two-dimensional histograms of the Monte Carlo simulations for $\Theta=0$ {\bf(a)}, $\pi/2$ {\bf(b)} and $\pi$ {\bf(c)}. Best agreement, \textit{i.e.} the maximum likelihood is found for $\Theta=\pi$. The example is taken for $\Omega t \approx \pi/2$ where the sensitivity is the highest.}

\end{figure*}

To compute the evolution of a given initial state under spin rotation, we use a Monte Carlo approach. The initial density operator is sampled by drawing random numbers $(n_0,m_z,\alpha)$ according to our assumed probability distributions (see Figure\,\ref{fig:InitialMLE}) and assuming a certain value for $\Theta$. This determines an initial mean field state $\vert \bm{\zeta}^N\rangle$. Using the known evolution under spin rotations, we propagate this state in time for a given $t_i$ to arrive at the final mean outcome populations $(n'_0,m'_z)$ as the expectation values of the corresponding operators in the time-evolved mean field state. In our numerical implementation we use a typical number of $\sim 10^6$ Monte Carlo samples to reconstruct the final statistical distribution of the measurement outcomes. Spin mixing collisions as discussed in Section\,\ref{sec:MFdynamics} are also taken into account to obtain the final simulated distributions. In the Monte-Carlo simulation, the spin state found after rotation is used as initial condition to solve the mean field equations (\ref{eq:spinmixing}) describing the spin dynamics. We arrive in this way at a distribution of $n'_0$ corrected for the effect of spin changing collisions, typically by a few percents. 

We evaluate the final populations for each realization using expectation values. Doing so, we neglect the effect of quantum fluctuations on the final results, which are on the order $1/\sqrt{N_{m_F}}$ and small for our typical atom numbers of particles ($N_{m_F}\sim$ a few thousands) when compared to the noise level of our population measurements. The measurement noise, caused by a combination of photon shot noise and small spatial intensity fluctuations of the laser pulse used for absorption imaging, is typically $\Delta n_{m_F}\approx 1 \%$ for the normalized population in Zeeman state $m_F$. We include this noise in our model by convolving the simulated measurement outcome by a Gaussian distribution. This leads to a conditional probability density $p_{t_i}(n'_0,m'_z |\Theta)$ for the measurement outcome which depends on the initial phase $\Theta$, which is then multiplied by the distribution $P(\Theta)$ to obtain $p_{t_i}(n'_0,m'_z |\bar{\Theta},\sigma_\Theta)$.

\subsection{Results of the MLE}
\label{sec:MLEdata}

\begin{figure*}[htb]
\begin{tabular}{lcr}
\centering
\includegraphics[width=0.33\linewidth]{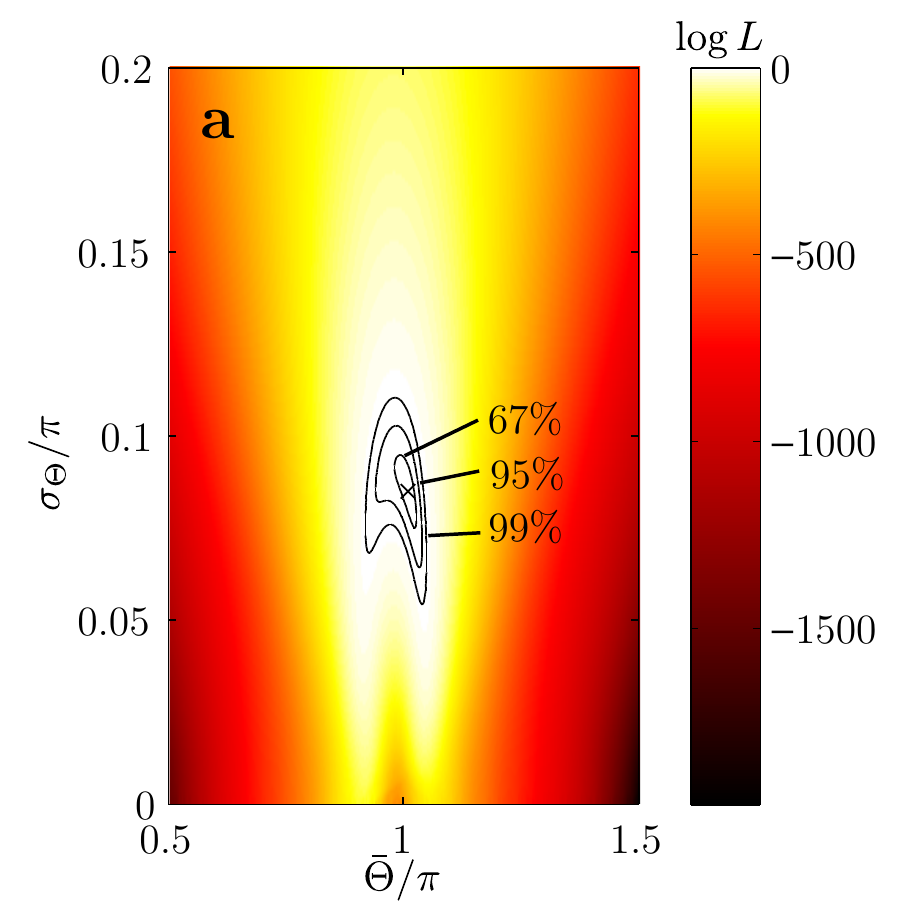}&
\includegraphics[width=0.33\linewidth]{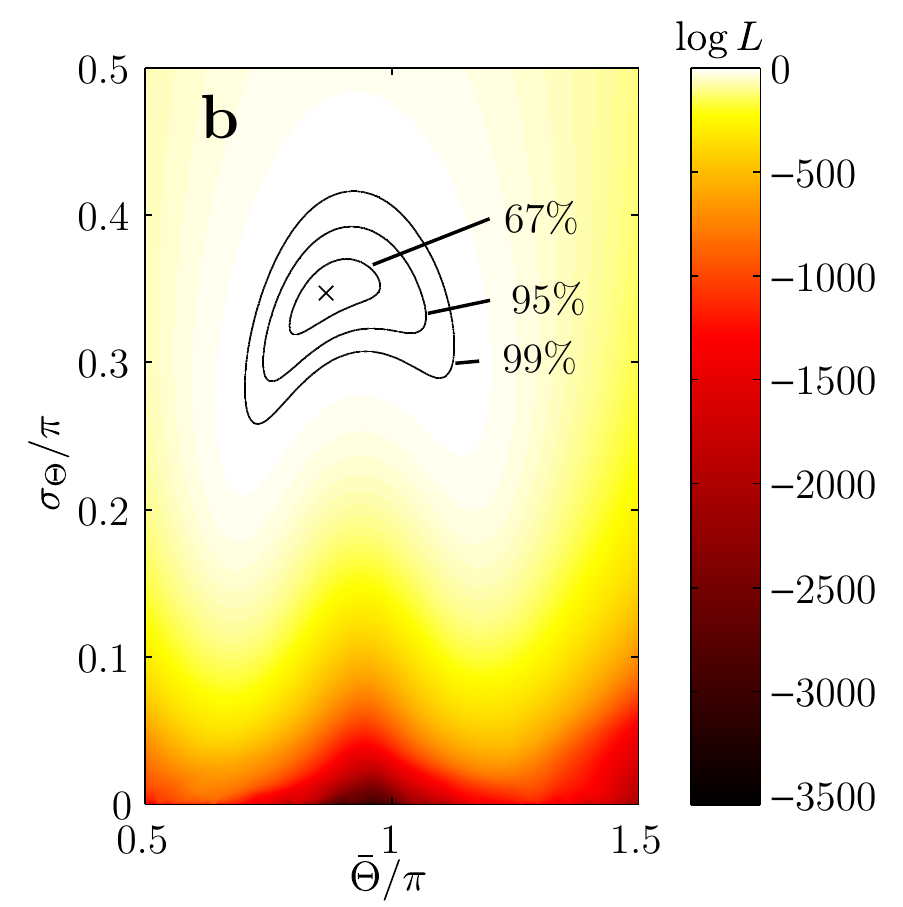}&
\includegraphics[width=0.33\linewidth]{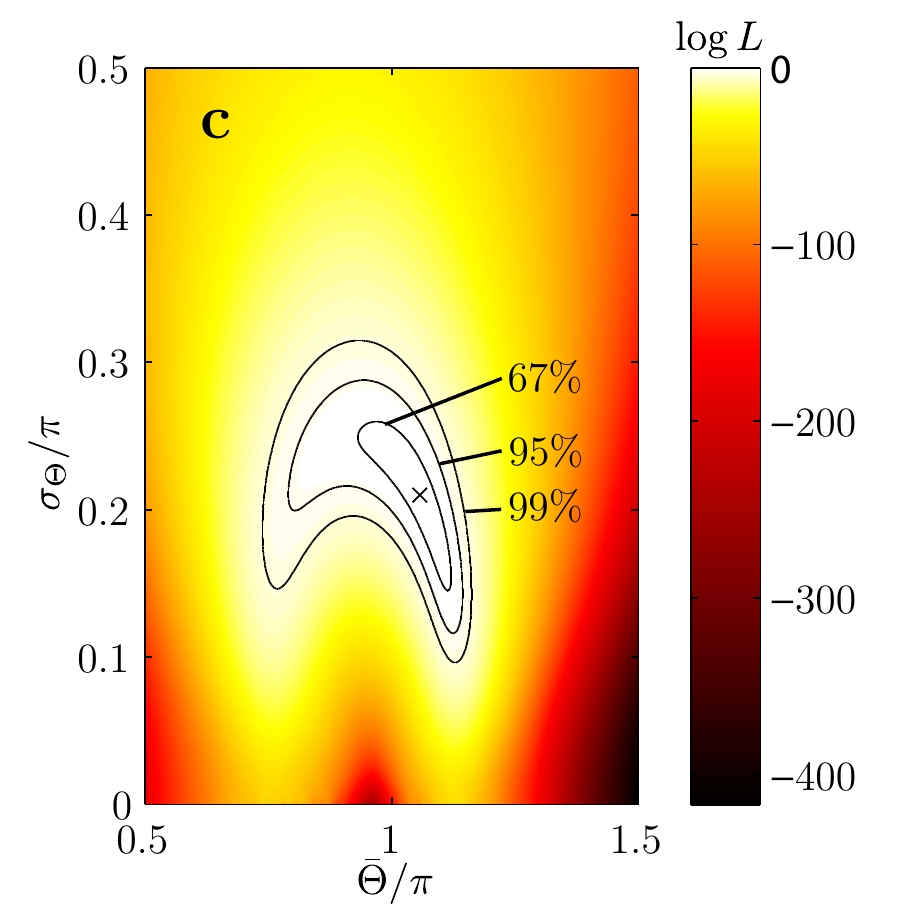}
\end{tabular}

\caption{(Color online) Results of the maximum likelihood determination of $P(\Theta)$, expressed in terms of the mean value $\bar{\Theta}$ and standard deviation $\sigma_\Theta$ for the datasets \textit{a}, \textit{b} and \textit{c}. In all cases the maximum, \textit{i.e.} the estimated phase $\Theta$, is found close to the theoretical predicted value of $\pi$ (black cross). The contour lines delineate the 67\%, 95\% and 99\% confidence area.} 
\label{fig:MLEAllResults}
\end{figure*}

\begin{figure}[t]
\centering
\includegraphics[trim=0cm 0cm 0cm 0cm, clip=true,width=0.5\textwidth]{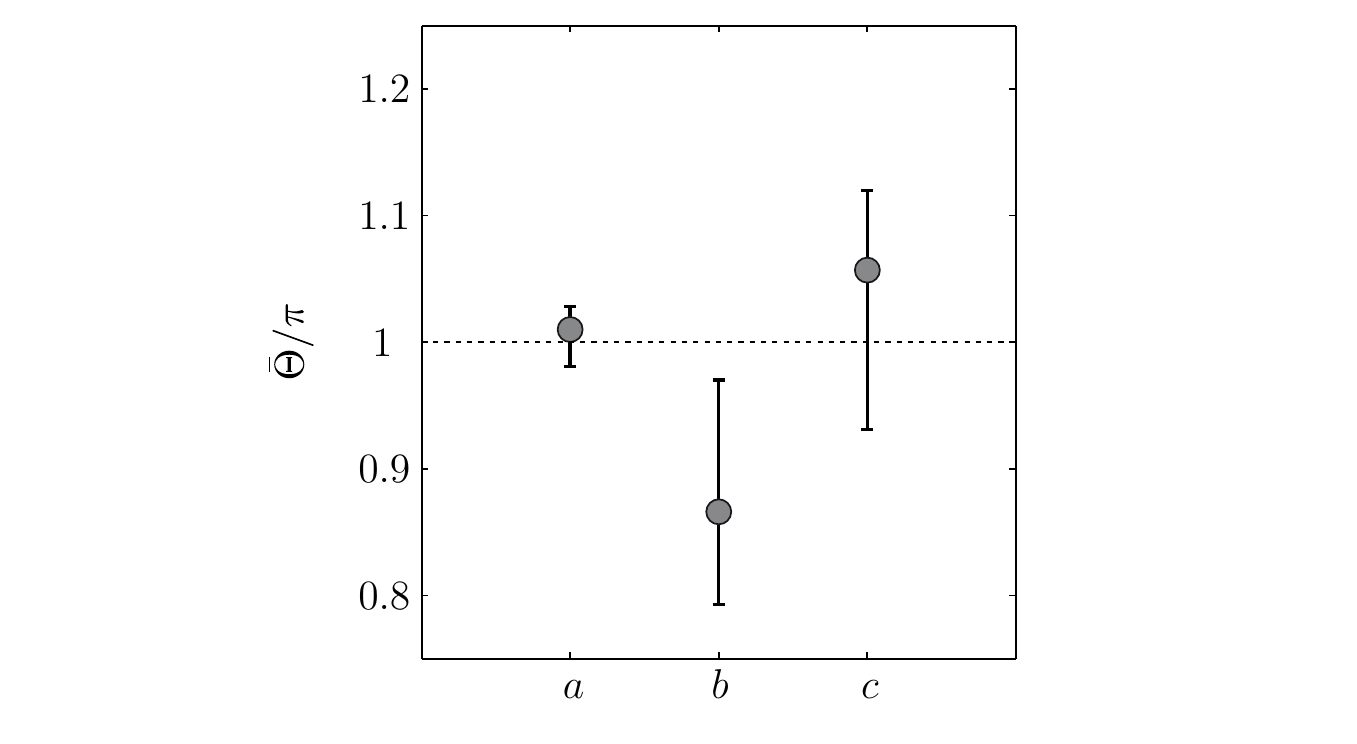}

\caption{(Color online) Most likely values of $\bar{\Theta}$ found by the MLE algorithm for data set \textit{a}, \textit{b}, \textit{c}. Error bars indicate the 67\% confidence bounds.}
\label{Fig:MLEResults}
\end{figure}

We model the distribution $P(\Theta)$ by a truncated Gaussian with a mean value $\bar{\Theta}$ and a standard deviation $\sigma_\Theta$. For the three data sets \textit{a}, \textit{b} and \textit{c}, the log likelihood is shown in Fig.\,\ref{fig:MLEAllResults} versus $(\bar{\Theta},\sigma_\Theta)$. The maxima, shown in Fig.\,\ref{Fig:MLEResults}, are found for $(\bar{\Theta},\sigma_\Theta)=(1.01\pi,0.085\pi)$, $(0.86\pi,0.347\pi)$ and $(1.05\pi,0.210\pi)$ for data sets \textit{a}, \textit{b} and \textit{c}, respectively. These results are in full agreement with the conclusion drawn from the variance analysis, confirming the locking to $\pi$ of the relative phase $\Theta$. 

In all instances, the MLE is maximum for a finite width $P(\Theta)$ which is not expected in the standard $T=0$ description. We conclude this Section by discussing possible explanations. First, it may be caused by an underestimation of the noise sources in the system. As seen before, the probability distribution $p_{t_i}(n'_0,m'_z |\Theta)$ is almost symmetric in $\Theta$ with respect to $0$ and $\pi$. The presence of fluctuations (induced for example by experimental imperfections) not included in our model always bias the estimator away from $\Theta=\pi$. We thus infer that underestimated or unconsidered noise in our probabilistic model will result in a broadening of the estimated distribution $P(\Theta)$. A second, more fundamental effect comes from the finite temperature of the initial spin ensemble (see Section\,\ref{sec:finiteT}). The marginal distribution of $\Theta$ obtained numerically \cite{corre2015a} is a bell-shape curve centered at $\pi$, reasonably approximated by a Gaussian with root-mean-square (rms) width $\approx \sqrt{k_B T/NU_s}$. Using $T\approx 80$\,nK and the experimental parameters of data set \textit{a}, we find a width $\approx 0.1$ comparable to the results of the MLE. 

\section{Conclusion}
\label{sec:conclusion}
In conclusion, we have shown the existence of spin-nematic ordering in antiferromagnetic spin 1 BECs, or equivalently of a phase locking between the Zeeman components caused by spin-exchange interactions in the equilibrium state. Our experimental method combines spin rotations with a statistical analysis, either based on the spin moments or on a maximum-likelihood estimation of the probability density function characterizing the initial spin state of the condensate. Our method is not restricted to single-mode condensates or to spin 1 atoms, and could be used to reveal other types of spin ordering. We remark in particular that measuring the spin variance gives access to a quantity (the squared transverse spin length) which can be used to characterize other phases than a fully condensed state. The expression of the transverse spin operator in Eq.\,(\ref{eq:Sx2Sy2}) shows that measuring the spin variance gives access to the ``spin singlet amplitude" $\left\langle \hat{a}_{+1}^\dagger\hat{a}_{+1}^\dagger \hat{a}_0^2 \right\rangle$ \cite{Law1998a,Koashi2000}, which appears in studies of fluctuating systems beyond mean field (spin liquid in one dimension \cite{Essler2009}, or spin-singlet Mott states in optical lattices, for instance \cite{zhou2003a}). 

\begin{acknowledgments}
We acknowledge support from IFRAF, from DARPA (OLE program), from the Hamburg Center for Ultrafast Imaging and from the ERC (Synergy grant UQUAM).
\end{acknowledgments}

\appendix

\section{Calculation of spin-mixing dynamics}\label{app:spinmixing}
To quantify the impact of spin-mixing oscillations on the measured $n_0$, we use the theoretical framework given in \cite{zhang2005a}. The evolution of an initial state of the form given in Eq.\,(\ref{eq:Psi}) is described by the two Josephson-like equations \cite{zhang2005a} ,
\begin{align}\label{eq:spinmixing}
\hbar\frac{d n_0}{dt}&= 2 U_s n_0 \sqrt{(1-n_0)^2-m_z^2}\sin(\Theta),\\
\hbar\frac{d \Theta}{dt}&= -2 q(t)+ 2 U_s(1-2n_0)\\
\nonumber &=+2U_s \frac{(1-n_0)(1-2n_0)-m_z^2}{\sqrt{(1-n_0)^2-m_z^2}}\cos(\Theta),
\end{align}
with $n_0(0)=n_{0,i}$ and $\Theta(0)=\Theta_i$. We solve Eqs.\,(\ref{eq:spinmixing}) numerically with $q(t)$ as shown in Fig.\,\ref{fig:MFdynamics}b to compute the evolution of $n_0$. 

\section{Hartree-Fock model of a spin 1 gas at finite-temperatures}
\label{app:HF}

The model of \cite{kawaguchi2012b} treats the non-condensed cloud as a gas of non-interacting free particles evolving in a self-consistent mean field potential accounting for spin-exchange interactions \cite{kawaguchi2012b}. Importantly, this mean field potential is not diagonal in the Zeeman basis due to spin-mixing interactions. The thermal component can in principle develop non-zero coherences due to interactions with the condensate and therefore a non-zero average spin. The quantity of interest is the single-particle density matrix,
\begin{align}
\rho_{m,n}^{(1)} (\bm{r})& = \phi_m^\ast(\bm{r}) \phi_n(\bm{r}) + \rho'^{(1)}_{m,n} (\bm{r}),
\end{align}
with $\bm{\phi}$ the condensate wavefunction, with $\rho'^{(1)}_{m,n}$ the contribution of the thermal component, and where $m,n=0,\pm1$. The
density in each Zeeman component $m$ is determined by the diagonal terms $\rho_{m,m}^{(1)}$ and the transverse spin by the off-diagonal coherences $\rho_{0,\pm 1}^{(1)}$.
 
With respect to the full HF model laid out in \cite{kawaguchi2012b}, we make two additional simplifying assumptions. First, we assume that the single-mode approximation holds for the condensate wavefunction \footnote{This was verified in an independent calculation by solving the three-component, three-dimensional Gross-Pitaevskii (GP) equation}. This amounts to setting $\phi_m(\bm{r})=\sqrt{N_c}\,\overline{\phi}(\bm{r} )\zeta_m$, as done in the main text. The single mode wavefunction $\overline{\phi}$ determining the condensate spatial distribution is computed numerically by solving the GP equation
\begin{align}
\label{eq:GPSMA}
\mu \overline{\phi} & =  - \frac{\hbar^2}{2 M_{\rm Na}} \Delta \overline{\phi} + V(\bm{r})  \overline{\phi}+ \overline{g} N_c\vert \overline{\phi}\vert^2 \overline{\phi}.
\end{align}
with $M_{\rm Na}$ the mass of Sodium atoms. The spinor part $\zeta_m$ is found from the single-mode theory using $U_s = N_c g_s \int d^{(3)} \bm{r} \vert \overline{\phi} \vert^4$. The coupling constants $\overline{g},g_s$ are proportional to the scattering lengths $\overline{a}\approx 2.79\,$nm and $a_s \approx 0.1\,$nm \cite{knoop2011a} with a proportionality factor $4\pi\hbar^2/M_{\rm Na}$.
Second, we neglect the contribution of the thermal cloud to the mean-field potential (``semi-ideal'' model \cite{naraschewski1998a}). Far from $T_c$, this is expected to be an accurate approximation \cite{dalfovo1999a}. Finally, we perform the calculations for a spherical trap. Although the trapping potential used in the experiment is not exactly isotropic, we do not expect that this affects strongly the results (in the Thomas-Fermi regime, for instance, only the average trap frequency matters to compute thermodynamic quantities \cite{dalfovo1999a}).

The excitations modes $\bm{u}^{(\nu)}$ and energies $E_\nu$ are solutions of the eigenproblem
\begin{align}
\label{eq:HFmodes}
E_\nu \bm{u}^{(\nu)}& = \left(- \frac{\hbar^2}{2M_{\rm Na}}\Delta \cdot + V(\bm{r}) + A(\bm{r}) \right)  \bm{u}^{(\nu)}
\end{align}
where the matrix $A$, explicitely  given in \cite{kawaguchi2012b}, depends on the condensate wavefunction $\bm{\phi}(\bm{r})$ and on $\overline{g},g_s$. Diagonalizing this equation, we obtain the single-particle density matrix $\rho'^{(1)}$ of the thermal component as
\begin{align}
\rho'^{(1)}_{m,n} (\bm{r} )  & = \sum_\nu \left( \bm{u}^{(\nu)}_m(\bm{r}) \right)^\ast \bm{u}^{(\nu)}_n(\bm{r}) N_{\rm BE}(E_\nu)
\end{align}
with $N_{\rm BE}(E)=1/(e^{E/k_B T}-1)$ the occupation number for each mode $\nu$.

\bibliography{BibSpinor}

\begin{thebibliography}{45}
\expandafter\ifx\csname natexlab\endcsname\relax\def\natexlab#1{#1}\fi
\expandafter\ifx\csname bibnamefont\endcsname\relax
  \def\bibnamefont#1{#1}\fi
\expandafter\ifx\csname bibfnamefont\endcsname\relax
  \def\bibfnamefont#1{#1}\fi
\expandafter\ifx\csname citenamefont\endcsname\relax
  \def\citenamefont#1{#1}\fi
\expandafter\ifx\csname url\endcsname\relax
  \def\url#1{\texttt{#1}}\fi
\expandafter\ifx\csname urlprefix\endcsname\relax\def\urlprefix{URL }\fi
\providecommand{\bibinfo}[2]{#2}
\providecommand{\eprint}[2][]{\url{#2}}

\bibitem[{\citenamefont{A.F.Andreev and I.A.Grishchuk}(1984)}]{andreev1984a}
\bibinfo{author}{\bibnamefont{A.F.Andreev}} \bibnamefont{and}
  \bibinfo{author}{\bibnamefont{I.A.Grishchuk}},
  \bibinfo{journal}{Zh.Eksp.Teor.Fiz.} \textbf{\bibinfo{volume}{87}},
  \bibinfo{pages}{467} (\bibinfo{year}{1984}), \bibinfo{note}{(Sov.Phys.-JETP
  1984, 60, pp. 267-271)}.

\bibitem[{\citenamefont{de~Gennes and Prost}(1995)}]{degennes}
\bibinfo{author}{\bibfnamefont{P.~G.} \bibnamefont{de~Gennes}}
  \bibnamefont{and} \bibinfo{author}{\bibfnamefont{J.}~\bibnamefont{Prost}},
  \emph{\bibinfo{title}{The Physics of Liquid Crystals}}
  (\bibinfo{publisher}{Clarendon Press}, \bibinfo{address}{Oxford},
  \bibinfo{year}{1995}).

\bibitem[{\citenamefont{Blume and Hsieh}(1969)}]{blume1969a}
\bibinfo{author}{\bibfnamefont{M.}~\bibnamefont{Blume}} \bibnamefont{and}
  \bibinfo{author}{\bibfnamefont{Y.~Y.} \bibnamefont{Hsieh}},
  \bibinfo{journal}{Journal of Applied Physics} \textbf{\bibinfo{volume}{40}},
  \bibinfo{pages}{1249} (\bibinfo{year}{1969}).

\bibitem[{\citenamefont{Chen and Levy}(1971)}]{chen1971a}
\bibinfo{author}{\bibfnamefont{H.~H.} \bibnamefont{Chen}} \bibnamefont{and}
  \bibinfo{author}{\bibfnamefont{P.~M.} \bibnamefont{Levy}},
  \bibinfo{journal}{Phys. Rev. Lett.} \textbf{\bibinfo{volume}{27}},
  \bibinfo{pages}{1383} (\bibinfo{year}{1971}).

\bibitem[{\citenamefont{Nakatsuji et~al.}(2005)\citenamefont{Nakatsuji, Nambu,
  Tonomura, Sakai, Jonas, Broholm, Tsunetsugu, Qiu, and
  Maeno}}]{Nakatsuji2005a}
\bibinfo{author}{\bibfnamefont{S.}~\bibnamefont{Nakatsuji}},
  \bibinfo{author}{\bibfnamefont{Y.}~\bibnamefont{Nambu}},
  \bibinfo{author}{\bibfnamefont{H.}~\bibnamefont{Tonomura}},
  \bibinfo{author}{\bibfnamefont{O.}~\bibnamefont{Sakai}},
  \bibinfo{author}{\bibfnamefont{S.}~\bibnamefont{Jonas}},
  \bibinfo{author}{\bibfnamefont{C.}~\bibnamefont{Broholm}},
  \bibinfo{author}{\bibfnamefont{H.}~\bibnamefont{Tsunetsugu}},
  \bibinfo{author}{\bibfnamefont{Y.}~\bibnamefont{Qiu}}, \bibnamefont{and}
  \bibinfo{author}{\bibfnamefont{Y.}~\bibnamefont{Maeno}},
  \bibinfo{journal}{Science} \textbf{\bibinfo{volume}{309}},
  \bibinfo{pages}{1697} (\bibinfo{year}{2005}).

\bibitem[{\citenamefont{Podolsky and Demler}(2005)}]{podolski2005a}
\bibinfo{author}{\bibfnamefont{D.}~\bibnamefont{Podolsky}} \bibnamefont{and}
  \bibinfo{author}{\bibfnamefont{E.}~\bibnamefont{Demler}},
  \bibinfo{journal}{New Journal of Physics} \textbf{\bibinfo{volume}{7}},
  \bibinfo{pages}{59} (\bibinfo{year}{2005}).

\bibitem[{\citenamefont{Tsunetsugu and Arikawa}(2006)}]{tsunetsugu2006a}
\bibinfo{author}{\bibfnamefont{H.}~\bibnamefont{Tsunetsugu}} \bibnamefont{and}
  \bibinfo{author}{\bibfnamefont{M.}~\bibnamefont{Arikawa}},
  \bibinfo{journal}{J. Phys. Soc. Jpn} \textbf{\bibinfo{volume}{75}},
  \bibinfo{pages}{083701} (\bibinfo{year}{2006}).

\bibitem[{\citenamefont{Bhattacharjee et~al.}(2006)\citenamefont{Bhattacharjee,
  Shenoy, and Senthil}}]{bhattacharjee2006a}
\bibinfo{author}{\bibfnamefont{S.}~\bibnamefont{Bhattacharjee}},
  \bibinfo{author}{\bibfnamefont{V.~B.} \bibnamefont{Shenoy}},
  \bibnamefont{and} \bibinfo{author}{\bibfnamefont{T.}~\bibnamefont{Senthil}},
  \bibinfo{journal}{Phys. Rev. B} \textbf{\bibinfo{volume}{74}},
  \bibinfo{pages}{092406} (\bibinfo{year}{2006}).

\bibitem[{\citenamefont{L\"auchli et~al.}(2006)\citenamefont{L\"auchli, Mila,
  and Penc}}]{lauchli2006a}
\bibinfo{author}{\bibfnamefont{A.}~\bibnamefont{L\"auchli}},
  \bibinfo{author}{\bibfnamefont{F.}~\bibnamefont{Mila}}, \bibnamefont{and}
  \bibinfo{author}{\bibfnamefont{K.}~\bibnamefont{Penc}},
  \bibinfo{journal}{Phys. Rev. Lett.} \textbf{\bibinfo{volume}{97}},
  \bibinfo{pages}{087205} (\bibinfo{year}{2006}).

\bibitem[{\citenamefont{Michaud et~al.}(2011)\citenamefont{Michaud, Vernay, and
  Mila}}]{michaud2011a}
\bibinfo{author}{\bibfnamefont{F.}~\bibnamefont{Michaud}},
  \bibinfo{author}{\bibfnamefont{F.}~\bibnamefont{Vernay}}, \bibnamefont{and}
  \bibinfo{author}{\bibfnamefont{F.}~\bibnamefont{Mila}},
  \bibinfo{journal}{Phys. Rev. B} \textbf{\bibinfo{volume}{84}},
  \bibinfo{pages}{184424} (\bibinfo{year}{2011}).

\bibitem[{\citenamefont{Stamper-Kurn and Ueda}(2013)}]{stamperkurn2013a}
\bibinfo{author}{\bibfnamefont{D.~M.} \bibnamefont{Stamper-Kurn}}
  \bibnamefont{and} \bibinfo{author}{\bibfnamefont{M.}~\bibnamefont{Ueda}},
  \bibinfo{journal}{Rev. Mod. Phys.} \textbf{\bibinfo{volume}{85}},
  \bibinfo{pages}{1191} (\bibinfo{year}{2013}).

\bibitem[{\citenamefont{Stenger et~al.}(1998)\citenamefont{Stenger, Inouye,
  Stamper-Kurn, Miesner, Chikkatur, and Ketterle}}]{stenger1998a}
\bibinfo{author}{\bibfnamefont{J.}~\bibnamefont{Stenger}},
  \bibinfo{author}{\bibfnamefont{S.}~\bibnamefont{Inouye}},
  \bibinfo{author}{\bibfnamefont{D.}~\bibnamefont{Stamper-Kurn}},
  \bibinfo{author}{\bibfnamefont{H.-J.} \bibnamefont{Miesner}},
  \bibinfo{author}{\bibfnamefont{A.}~\bibnamefont{Chikkatur}},
  \bibnamefont{and} \bibinfo{author}{\bibfnamefont{W.}~\bibnamefont{Ketterle}},
  \bibinfo{journal}{Nature} \textbf{\bibinfo{volume}{396}},
  \bibinfo{pages}{345} (\bibinfo{year}{1998}).

\bibitem[{\citenamefont{Ohmi and Machida}(1998)}]{ohmi1998a}
\bibinfo{author}{\bibfnamefont{T.}~\bibnamefont{Ohmi}} \bibnamefont{and}
  \bibinfo{author}{\bibfnamefont{T.}~\bibnamefont{Machida}},
  \bibinfo{journal}{J. Phys. Soc. Jpn} \textbf{\bibinfo{volume}{67}},
  \bibinfo{pages}{1822} (\bibinfo{year}{1998}).

\bibitem[{\citenamefont{Snoek and Zhou}(2004)}]{zhou2003a}
\bibinfo{author}{\bibfnamefont{M.}~\bibnamefont{Snoek}} \bibnamefont{and}
  \bibinfo{author}{\bibfnamefont{F.}~\bibnamefont{Zhou}},
  \bibinfo{journal}{Phys. Rev. B} \textbf{\bibinfo{volume}{69}},
  \bibinfo{pages}{094410} (\bibinfo{year}{2004}).

\bibitem[{\citenamefont{Imambekov et~al.}(2003)\citenamefont{Imambekov, Lukin,
  and Demler}}]{imambekov2003a}
\bibinfo{author}{\bibfnamefont{A.}~\bibnamefont{Imambekov}},
  \bibinfo{author}{\bibfnamefont{M.~D.} \bibnamefont{Lukin}}, \bibnamefont{and}
  \bibinfo{author}{\bibfnamefont{E.}~\bibnamefont{Demler}},
  \bibinfo{journal}{Phys. Rev. A} \textbf{\bibinfo{volume}{68}},
  \bibinfo{pages}{063602} (\bibinfo{year}{2003}).

\bibitem[{\citenamefont{Zhou et~al.}(2004)\citenamefont{Zhou, Snoek, Wiemer,
  and Affleck}}]{zhou2004a}
\bibinfo{author}{\bibfnamefont{F.}~\bibnamefont{Zhou}},
  \bibinfo{author}{\bibfnamefont{M.}~\bibnamefont{Snoek}},
  \bibinfo{author}{\bibfnamefont{J.}~\bibnamefont{Wiemer}}, \bibnamefont{and}
  \bibinfo{author}{\bibfnamefont{I.}~\bibnamefont{Affleck}},
  \bibinfo{journal}{Phys. Rev. B} \textbf{\bibinfo{volume}{70}},
  \bibinfo{pages}{184434} (\bibinfo{year}{2004}).

\bibitem[{\citenamefont{Black et~al.}(2007)\citenamefont{Black, Gomez, Turner,
  Jung, and Lett}}]{black2007a}
\bibinfo{author}{\bibfnamefont{A.~T.} \bibnamefont{Black}},
  \bibinfo{author}{\bibfnamefont{E.}~\bibnamefont{Gomez}},
  \bibinfo{author}{\bibfnamefont{L.~D.} \bibnamefont{Turner}},
  \bibinfo{author}{\bibfnamefont{S.}~\bibnamefont{Jung}}, \bibnamefont{and}
  \bibinfo{author}{\bibfnamefont{P.~D.} \bibnamefont{Lett}},
  \bibinfo{journal}{Phys. Rev. Lett.} \textbf{\bibinfo{volume}{99}},
  \bibinfo{pages}{070403} (\bibinfo{year}{2007}).

\bibitem[{\citenamefont{Liu et~al.}(2009)\citenamefont{Liu, Jung, Maxwell,
  Turner, Tiesinga, and Lett}}]{liu2009a}
\bibinfo{author}{\bibfnamefont{Y.}~\bibnamefont{Liu}},
  \bibinfo{author}{\bibfnamefont{S.}~\bibnamefont{Jung}},
  \bibinfo{author}{\bibfnamefont{S.~E.} \bibnamefont{Maxwell}},
  \bibinfo{author}{\bibfnamefont{L.~D.} \bibnamefont{Turner}},
  \bibinfo{author}{\bibfnamefont{E.}~\bibnamefont{Tiesinga}}, \bibnamefont{and}
  \bibinfo{author}{\bibfnamefont{P.~D.} \bibnamefont{Lett}},
  \bibinfo{journal}{Phys. Rev. Lett.} \textbf{\bibinfo{volume}{102}},
  \bibinfo{pages}{125301} (\bibinfo{year}{2009}).

\bibitem[{\citenamefont{Bookjans et~al.}(2011)\citenamefont{Bookjans, Vinit,
  and Raman}}]{bookjans2011a}
\bibinfo{author}{\bibfnamefont{E.~M.} \bibnamefont{Bookjans}},
  \bibinfo{author}{\bibfnamefont{A.}~\bibnamefont{Vinit}}, \bibnamefont{and}
  \bibinfo{author}{\bibfnamefont{C.}~\bibnamefont{Raman}},
  \bibinfo{journal}{Phys. Rev. Lett.} \textbf{\bibinfo{volume}{107}},
  \bibinfo{pages}{195306} (\bibinfo{year}{2011}).

\bibitem[{\citenamefont{Jacob et~al.}(2012)\citenamefont{Jacob, Shao, Corre,
  Zibold, De~Sarlo, Mimoun, Dalibard, and Gerbier}}]{jacob2012a}
\bibinfo{author}{\bibfnamefont{D.}~\bibnamefont{Jacob}},
  \bibinfo{author}{\bibfnamefont{L.}~\bibnamefont{Shao}},
  \bibinfo{author}{\bibfnamefont{V.}~\bibnamefont{Corre}},
  \bibinfo{author}{\bibfnamefont{T.}~\bibnamefont{Zibold}},
  \bibinfo{author}{\bibfnamefont{L.}~\bibnamefont{De~Sarlo}},
  \bibinfo{author}{\bibfnamefont{E.}~\bibnamefont{Mimoun}},
  \bibinfo{author}{\bibfnamefont{J.}~\bibnamefont{Dalibard}}, \bibnamefont{and}
  \bibinfo{author}{\bibfnamefont{F.}~\bibnamefont{Gerbier}},
  \bibinfo{journal}{Phys. Rev. A} \textbf{\bibinfo{volume}{86}},
  \bibinfo{pages}{061601} (\bibinfo{year}{2012}).

\bibitem[{\citenamefont{de~Forges~de Parny
  et~al.}(2014)\citenamefont{de~Forges~de Parny, Yang, and
  Mila}}]{deforges2014a}
\bibinfo{author}{\bibfnamefont{L.}~\bibnamefont{de~Forges~de Parny}},
  \bibinfo{author}{\bibfnamefont{H.}~\bibnamefont{Yang}}, \bibnamefont{and}
  \bibinfo{author}{\bibfnamefont{F.}~\bibnamefont{Mila}},
  \bibinfo{journal}{Phys. Rev. Lett.} \textbf{\bibinfo{volume}{113}},
  \bibinfo{pages}{200402} (\bibinfo{year}{2014}).

\bibitem[{\citenamefont{Kawaguchi and Ueda}(2012)}]{ueda2012a}
\bibinfo{author}{\bibfnamefont{Y.}~\bibnamefont{Kawaguchi}} \bibnamefont{and}
  \bibinfo{author}{\bibfnamefont{M.}~\bibnamefont{Ueda}},
  \bibinfo{journal}{Physics Reports} \textbf{\bibinfo{volume}{520}},
  \bibinfo{pages}{253 } (\bibinfo{year}{2012}).

\bibitem[{\citenamefont{Yi et~al.}(2002)\citenamefont{Yi, M\"ustecapl\ifmmode
  \imath \else \i \fi{}o\ifmmode~\breve{g}\else \u{g}\fi{}lu, Sun, and
  You}}]{yi2002a}
\bibinfo{author}{\bibfnamefont{S.}~\bibnamefont{Yi}},
  \bibinfo{author}{\bibfnamefont{O.~E.} \bibnamefont{M\"ustecapl\ifmmode \imath
  \else \i \fi{}o\ifmmode~\breve{g}\else \u{g}\fi{}lu}},
  \bibinfo{author}{\bibfnamefont{C.~P.} \bibnamefont{Sun}}, \bibnamefont{and}
  \bibinfo{author}{\bibfnamefont{L.}~\bibnamefont{You}},
  \bibinfo{journal}{Phys. Rev. A} \textbf{\bibinfo{volume}{66}},
  \bibinfo{pages}{011601} (\bibinfo{year}{2002}).

\bibitem[{\citenamefont{Mullin et~al.}(1966)\citenamefont{Mullin, Keller,
  Hammer, and Good}}]{mullin1966a}
\bibinfo{author}{\bibfnamefont{C.~J.} \bibnamefont{Mullin}},
  \bibinfo{author}{\bibfnamefont{J.~M.} \bibnamefont{Keller}},
  \bibinfo{author}{\bibfnamefont{C.~L.} \bibnamefont{Hammer}},
  \bibnamefont{and} \bibinfo{author}{\bibfnamefont{R.~H.} \bibnamefont{Good}},
  \bibinfo{journal}{Annals of Physics} \textbf{\bibinfo{volume}{37}},
  \bibinfo{pages}{55} (\bibinfo{year}{1966}).

\bibitem[{\citenamefont{Ivanov and Kolezhuk}(2003)}]{ivanov2003a}
\bibinfo{author}{\bibfnamefont{B.~A.} \bibnamefont{Ivanov}} \bibnamefont{and}
  \bibinfo{author}{\bibfnamefont{A.~K.} \bibnamefont{Kolezhuk}},
  \bibinfo{journal}{Phys. Rev. B} \textbf{\bibinfo{volume}{68}},
  \bibinfo{pages}{052401} (\bibinfo{year}{2003}).

\bibitem[{\citenamefont{Zhang et~al.}(2003)\citenamefont{Zhang, Yi, and
  You}}]{zhang2003a}
\bibinfo{author}{\bibfnamefont{W.}~\bibnamefont{Zhang}},
  \bibinfo{author}{\bibfnamefont{S.}~\bibnamefont{Yi}}, \bibnamefont{and}
  \bibinfo{author}{\bibfnamefont{L.}~\bibnamefont{You}}, \bibinfo{journal}{New
  Journal of Physics} \textbf{\bibinfo{volume}{5}}, \bibinfo{pages}{77}
  (\bibinfo{year}{2003}).

\bibitem[{\citenamefont{Corre et~al.}(2015)\citenamefont{Corre, Zibold,
  Frapolli, Shao, Dalibard, and Gerbier}}]{corre2015a}
\bibinfo{author}{\bibfnamefont{V.}~\bibnamefont{Corre}},
  \bibinfo{author}{\bibfnamefont{T.}~\bibnamefont{Zibold}},
  \bibinfo{author}{\bibfnamefont{C.}~\bibnamefont{Frapolli}},
  \bibinfo{author}{\bibfnamefont{L.}~\bibnamefont{Shao}},
  \bibinfo{author}{\bibfnamefont{J.}~\bibnamefont{Dalibard}}, \bibnamefont{and}
  \bibinfo{author}{\bibfnamefont{F.}~\bibnamefont{Gerbier}},
  \bibinfo{journal}{EPL} \textbf{\bibinfo{volume}{110}}, \bibinfo{pages}{26001}
  (\bibinfo{year}{2015}).

\bibitem[{\citenamefont{Kawaguchi et~al.}(2012)\citenamefont{Kawaguchi, Phuc,
  and Blakie}}]{kawaguchi2012b}
\bibinfo{author}{\bibfnamefont{Y.}~\bibnamefont{Kawaguchi}},
  \bibinfo{author}{\bibfnamefont{N.~T.} \bibnamefont{Phuc}}, \bibnamefont{and}
  \bibinfo{author}{\bibfnamefont{P.~B.} \bibnamefont{Blakie}},
  \bibinfo{journal}{Phys. Rev. A} \textbf{\bibinfo{volume}{85}},
  \bibinfo{pages}{053611} (\bibinfo{year}{2012}).

\bibitem[{\citenamefont{L{\"u}cke et~al.}(2011)\citenamefont{L{\"u}cke,
  Scherer, Kruse, Pezz{\'e}, Deuretzbacher, Hyllus, Peise, Ertmer, Arlt, Santos
  et~al.}}]{lucke2011twin}
\bibinfo{author}{\bibfnamefont{B.}~\bibnamefont{L{\"u}cke}},
  \bibinfo{author}{\bibfnamefont{M.}~\bibnamefont{Scherer}},
  \bibinfo{author}{\bibfnamefont{J.}~\bibnamefont{Kruse}},
  \bibinfo{author}{\bibfnamefont{L.}~\bibnamefont{Pezz{\'e}}},
  \bibinfo{author}{\bibfnamefont{F.}~\bibnamefont{Deuretzbacher}},
  \bibinfo{author}{\bibfnamefont{P.}~\bibnamefont{Hyllus}},
  \bibinfo{author}{\bibfnamefont{J.}~\bibnamefont{Peise}},
  \bibinfo{author}{\bibfnamefont{W.}~\bibnamefont{Ertmer}},
  \bibinfo{author}{\bibfnamefont{J.}~\bibnamefont{Arlt}},
  \bibinfo{author}{\bibfnamefont{L.}~\bibnamefont{Santos}},
  \bibnamefont{et~al.}, \bibinfo{journal}{Science}
  \textbf{\bibinfo{volume}{334}}, \bibinfo{pages}{773} (\bibinfo{year}{2011}).

\bibitem[{\citenamefont{Gross et~al.}(2011)\citenamefont{Gross, Strobel,
  Nicklas, Zibold, Bar-Gill, Kurizki, and Oberthaler}}]{gross2011atomic}
\bibinfo{author}{\bibfnamefont{C.}~\bibnamefont{Gross}},
  \bibinfo{author}{\bibfnamefont{H.}~\bibnamefont{Strobel}},
  \bibinfo{author}{\bibfnamefont{E.}~\bibnamefont{Nicklas}},
  \bibinfo{author}{\bibfnamefont{T.}~\bibnamefont{Zibold}},
  \bibinfo{author}{\bibfnamefont{N.}~\bibnamefont{Bar-Gill}},
  \bibinfo{author}{\bibfnamefont{G.}~\bibnamefont{Kurizki}}, \bibnamefont{and}
  \bibinfo{author}{\bibfnamefont{M.}~\bibnamefont{Oberthaler}},
  \bibinfo{journal}{Nature} \textbf{\bibinfo{volume}{480}},
  \bibinfo{pages}{219} (\bibinfo{year}{2011}).

\bibitem[{\citenamefont{Hamley et~al.}(2012)\citenamefont{Hamley, Gerving,
  Hoang, Bookjans, and Chapman}}]{hamley2012spin}
\bibinfo{author}{\bibfnamefont{C.}~\bibnamefont{Hamley}},
  \bibinfo{author}{\bibfnamefont{C.}~\bibnamefont{Gerving}},
  \bibinfo{author}{\bibfnamefont{T.}~\bibnamefont{Hoang}},
  \bibinfo{author}{\bibfnamefont{E.}~\bibnamefont{Bookjans}}, \bibnamefont{and}
  \bibinfo{author}{\bibfnamefont{M.}~\bibnamefont{Chapman}},
  \bibinfo{journal}{Nature Physics} \textbf{\bibinfo{volume}{8}},
  \bibinfo{pages}{305} (\bibinfo{year}{2012}).

\bibitem[{\citenamefont{L{\"u}cke et~al.}(2014)\citenamefont{L{\"u}cke, Peise,
  Vitagliano, Arlt, Santos, T{\'o}th, and Klempt}}]{lucke2014detecting}
\bibinfo{author}{\bibfnamefont{B.}~\bibnamefont{L{\"u}cke}},
  \bibinfo{author}{\bibfnamefont{J.}~\bibnamefont{Peise}},
  \bibinfo{author}{\bibfnamefont{G.}~\bibnamefont{Vitagliano}},
  \bibinfo{author}{\bibfnamefont{J.}~\bibnamefont{Arlt}},
  \bibinfo{author}{\bibfnamefont{L.}~\bibnamefont{Santos}},
  \bibinfo{author}{\bibfnamefont{G.}~\bibnamefont{T{\'o}th}}, \bibnamefont{and}
  \bibinfo{author}{\bibfnamefont{C.}~\bibnamefont{Klempt}},
  \bibinfo{journal}{Phys. Rev. Lett.} \textbf{\bibinfo{volume}{112}},
  \bibinfo{pages}{155304} (\bibinfo{year}{2014}).

\bibitem[{\citenamefont{Jacob et~al.}(2011)\citenamefont{Jacob, Mimoun, Sarlo,
  Weitz, Dalibard, and Gerbier}}]{jacob2011a}
\bibinfo{author}{\bibfnamefont{D.}~\bibnamefont{Jacob}},
  \bibinfo{author}{\bibfnamefont{E.}~\bibnamefont{Mimoun}},
  \bibinfo{author}{\bibfnamefont{L.~D.} \bibnamefont{Sarlo}},
  \bibinfo{author}{\bibfnamefont{M.}~\bibnamefont{Weitz}},
  \bibinfo{author}{\bibfnamefont{J.}~\bibnamefont{Dalibard}}, \bibnamefont{and}
  \bibinfo{author}{\bibfnamefont{F.}~\bibnamefont{Gerbier}},
  \bibinfo{journal}{New Journal of Physics} \textbf{\bibinfo{volume}{13}},
  \bibinfo{pages}{065022} (\bibinfo{year}{2011}).

\bibitem[{\citenamefont{Ketterle et~al.}(1999)\citenamefont{Ketterle, Durfee,
  and Stamper-Kurn}}]{ketterlereview}
\bibinfo{author}{\bibfnamefont{W.}~\bibnamefont{Ketterle}},
  \bibinfo{author}{\bibfnamefont{D.~S.} \bibnamefont{Durfee}},
  \bibnamefont{and} \bibinfo{author}{\bibfnamefont{D.~M.}
  \bibnamefont{Stamper-Kurn}}, in \emph{\bibinfo{booktitle}{Proceedings of the
  International School on Physics Enrico Fermi 1998, Bose-Einstein Condensation
  in Atomic Gases}}, edited by
  \bibinfo{editor}{\bibfnamefont{M.}~\bibnamefont{Inguscio}},
  \bibinfo{editor}{\bibfnamefont{S.}~\bibnamefont{Stringari}},
  \bibnamefont{and} \bibinfo{editor}{\bibfnamefont{C.~E.} \bibnamefont{Wieman}}
  (\bibinfo{publisher}{IOS Press}, \bibinfo{year}{1999}), pp.
  \bibinfo{pages}{67--176}, \bibinfo{note}{arXiv:cond-mat/9904034}.

\bibitem[{\citenamefont{Pu et~al.}(1999)\citenamefont{Pu, Law, Raghavan,
  Eberly, and Bigelow}}]{pu1999a}
\bibinfo{author}{\bibfnamefont{H.}~\bibnamefont{Pu}},
  \bibinfo{author}{\bibfnamefont{C.~K.} \bibnamefont{Law}},
  \bibinfo{author}{\bibfnamefont{S.}~\bibnamefont{Raghavan}},
  \bibinfo{author}{\bibfnamefont{J.~H.} \bibnamefont{Eberly}},
  \bibnamefont{and} \bibinfo{author}{\bibfnamefont{N.~P.}
  \bibnamefont{Bigelow}}, \bibinfo{journal}{Phys. Rev. A}
  \textbf{\bibinfo{volume}{60}}, \bibinfo{pages}{1463} (\bibinfo{year}{1999}).

\bibitem[{\citenamefont{Chang et~al.}(2005)\citenamefont{Chang, Qin, Zhang,
  You, and Chapman}}]{chang2005a}
\bibinfo{author}{\bibfnamefont{M.-S.} \bibnamefont{Chang}},
  \bibinfo{author}{\bibfnamefont{Q.}~\bibnamefont{Qin}},
  \bibinfo{author}{\bibfnamefont{W.}~\bibnamefont{Zhang}},
  \bibinfo{author}{\bibfnamefont{L.}~\bibnamefont{You}}, \bibnamefont{and}
  \bibinfo{author}{\bibfnamefont{M.~S.} \bibnamefont{Chapman}},
  \bibinfo{journal}{Nature Physics} \textbf{\bibinfo{volume}{1}},
  \bibinfo{pages}{111} (\bibinfo{year}{2005}).

\bibitem[{\citenamefont{Kronj{\"a}ger et~al.}(2005)\citenamefont{Kronj{\"a}ger,
  Becker, Brinkmann, Walser, Navez, Bongs, and Sengstock}}]{kronjaeger2005a}
\bibinfo{author}{\bibfnamefont{J.}~\bibnamefont{Kronj{\"a}ger}},
  \bibinfo{author}{\bibfnamefont{C.}~\bibnamefont{Becker}},
  \bibinfo{author}{\bibfnamefont{M.}~\bibnamefont{Brinkmann}},
  \bibinfo{author}{\bibfnamefont{R.}~\bibnamefont{Walser}},
  \bibinfo{author}{\bibfnamefont{P.}~\bibnamefont{Navez}},
  \bibinfo{author}{\bibfnamefont{K.}~\bibnamefont{Bongs}}, \bibnamefont{and}
  \bibinfo{author}{\bibfnamefont{K.}~\bibnamefont{Sengstock}},
  \bibinfo{journal}{Physical Review A} \textbf{\bibinfo{volume}{72}},
  \bibinfo{pages}{063619} (\bibinfo{year}{2005}).

\bibitem[{\citenamefont{Zhang et~al.}(2005)\citenamefont{Zhang, Zhou, Chang,
  Chapman, and You}}]{zhang2005a}
\bibinfo{author}{\bibfnamefont{W.}~\bibnamefont{Zhang}},
  \bibinfo{author}{\bibfnamefont{D.~L.} \bibnamefont{Zhou}},
  \bibinfo{author}{\bibfnamefont{M.-S.} \bibnamefont{Chang}},
  \bibinfo{author}{\bibfnamefont{M.~S.} \bibnamefont{Chapman}},
  \bibnamefont{and} \bibinfo{author}{\bibfnamefont{L.}~\bibnamefont{You}},
  \bibinfo{journal}{Phys. Rev. A} \textbf{\bibinfo{volume}{72}},
  \bibinfo{pages}{013602} (\bibinfo{year}{2005}).

\bibitem[{\citenamefont{Ueda}(2000)}]{ueda2001a}
\bibinfo{author}{\bibfnamefont{M.}~\bibnamefont{Ueda}}, \bibinfo{journal}{Phys.
  Rev. A} \textbf{\bibinfo{volume}{63}}, \bibinfo{pages}{013601}
  (\bibinfo{year}{2000}).

\bibitem[{\citenamefont{Law et~al.}(1998)\citenamefont{Law, Pu, and
  Bigelow}}]{Law1998a}
\bibinfo{author}{\bibfnamefont{C.~K.} \bibnamefont{Law}},
  \bibinfo{author}{\bibfnamefont{H.}~\bibnamefont{Pu}}, \bibnamefont{and}
  \bibinfo{author}{\bibfnamefont{N.~P.} \bibnamefont{Bigelow}},
  \bibinfo{journal}{Phys. Rev. Lett.} \textbf{\bibinfo{volume}{81}},
  \bibinfo{pages}{5257} (\bibinfo{year}{1998}).

\bibitem[{\citenamefont{Koashi and Ueda}(2000)}]{Koashi2000}
\bibinfo{author}{\bibfnamefont{M.}~\bibnamefont{Koashi}} \bibnamefont{and}
  \bibinfo{author}{\bibfnamefont{M.}~\bibnamefont{Ueda}},
  \bibinfo{journal}{Phys. Rev. Lett.} \textbf{\bibinfo{volume}{84}},
  \bibinfo{pages}{1066} (\bibinfo{year}{2000}).

\bibitem[{\citenamefont{Essler et~al.}(2009)\citenamefont{Essler, Shlyapnikov,
  and Tsvelik}}]{Essler2009}
\bibinfo{author}{\bibfnamefont{F.~H.~L.} \bibnamefont{Essler}},
  \bibinfo{author}{\bibfnamefont{G.~V.} \bibnamefont{Shlyapnikov}},
  \bibnamefont{and} \bibinfo{author}{\bibfnamefont{A.~M.}
  \bibnamefont{Tsvelik}}, \bibinfo{journal}{Journal of Statistical Mechanics}
  \textbf{\bibinfo{volume}{02}}, \bibinfo{pages}{P02027}
  (\bibinfo{year}{2009}).

\bibitem[{\citenamefont{Knoop et~al.}(2011)\citenamefont{Knoop, Schuster,
  Scelle, Trautmann, Appmeier, Oberthaler, Tiesinga, and Tiemann}}]{knoop2011a}
\bibinfo{author}{\bibfnamefont{S.}~\bibnamefont{Knoop}},
  \bibinfo{author}{\bibfnamefont{T.}~\bibnamefont{Schuster}},
  \bibinfo{author}{\bibfnamefont{R.}~\bibnamefont{Scelle}},
  \bibinfo{author}{\bibfnamefont{A.}~\bibnamefont{Trautmann}},
  \bibinfo{author}{\bibfnamefont{J.}~\bibnamefont{Appmeier}},
  \bibinfo{author}{\bibfnamefont{M.~K.} \bibnamefont{Oberthaler}},
  \bibinfo{author}{\bibfnamefont{E.}~\bibnamefont{Tiesinga}}, \bibnamefont{and}
  \bibinfo{author}{\bibfnamefont{E.}~\bibnamefont{Tiemann}},
  \bibinfo{journal}{Phys. Rev. A} \textbf{\bibinfo{volume}{83}},
  \bibinfo{pages}{042704} (\bibinfo{year}{2011}).

\bibitem[{\citenamefont{Naraschewski and
  Stamper-Kurn}(1998)}]{naraschewski1998a}
\bibinfo{author}{\bibfnamefont{M.}~\bibnamefont{Naraschewski}}
  \bibnamefont{and} \bibinfo{author}{\bibfnamefont{D.~M.}
  \bibnamefont{Stamper-Kurn}}, \bibinfo{journal}{Phys. Rev. A}
  \textbf{\bibinfo{volume}{58}}, \bibinfo{pages}{2423} (\bibinfo{year}{1998}).

\bibitem[{\citenamefont{Dalfovo et~al.}(1999)\citenamefont{Dalfovo, Giorgini,
  Pitaevskii, and Stringari}}]{dalfovo1999a}
\bibinfo{author}{\bibfnamefont{F.}~\bibnamefont{Dalfovo}},
  \bibinfo{author}{\bibfnamefont{S.}~\bibnamefont{Giorgini}},
  \bibinfo{author}{\bibfnamefont{L.~P.} \bibnamefont{Pitaevskii}},
  \bibnamefont{and}
  \bibinfo{author}{\bibfnamefont{S.}~\bibnamefont{Stringari}},
  \bibinfo{journal}{Rev. Mod. Phys.} \textbf{\bibinfo{volume}{71}},
  \bibinfo{pages}{463} (\bibinfo{year}{1999}).

\end{thebibliography}
\bibliographystyle{apsrev}

\end{document}